%% file: camera_ready.tex
\documentclass[sigconf]{acmart}
\usepackage{tikz}

\AtBeginDocument{%
  \providecommand\BibTeX{{%
    \normalfont B\kern-0.5em{\scshape i\kern-0.25em b}\kern-0.8em\TeX}}}

\setcopyright{acmcopyright}
\copyrightyear{2024}
\acmYear{2024}
\setcopyright{rightsretained}
\acmConference[DIS '24]{Designing Interactive Systems Conference}{July 1--5, 2024}{IT University of Copenhagen, Denmark}
\acmBooktitle{Designing Interactive Systems Conference (DIS '24), July 1--5, 2024, IT University of Copenhagen, Denmark}\acmDOI{10.1145/3643834.3661568}
\acmISBN{979-8-4007-0583-0/24/07}

\begin{document}

\title{Expanding the Design Space of Computer Vision-based Interactive Systems for Group Dance Practice}

\author{Soohwan Lee}
\affiliation{%
  \institution{Department of Design, UNIST}
  \city{Ulsan}
  \country{Republic of Korea}}
\email{soohwanlee@unist.ac.kr}

\author{Seoyeong Hwang}
\affiliation{%
  \institution{Department of Design, UNIST}
  \city{Ulsan}
  \country{Republic of Korea}}
\email{hseoyeong@unist.ac.kr}

\author{Ian Oakley}
\affiliation{%
  \institution{School of Electrical Engineering, KAIST}
  \city{Daejeon}
  \country{Republic of Korea}}
\email{ian.r.oakley@gmail.com}

\author{Kyungho Lee}
\affiliation{%
  \institution{Department of Design, UNIST}
  \city{Ulsan}
  \country{Republic of Korea}}
\email{kyungho@unist.ac.kr}

\renewcommand{\shortauthors}{Lee, et al.}
\newcommand{\KH}[1]{\textcolor{blue}{#1}}

\begin{abstract}
Group dance, a sub-genre characterized by intricate motions made by a cohort of performers in tight synchronization, has a longstanding and culturally significant history and, in modern forms such as cheerleading, a broad base of current adherents. However, despite its popularity, learning group dance routines remains challenging. Based on the prior success of interactive systems to support individual dance learning, this paper argues that group dance settings are fertile ground for augmentation by interactive aids. To better understand these design opportunities, this paper presents a sequence of user-centered studies of and with amateur cheerleading troupes, spanning from the formative (interviews, observations) through the generative (an ideation workshop) to concept validation (technology probes and speed dating). The outcomes are a nuanced understanding of the lived practice of group dance learning, a set of interactive concepts to support those practices, and design directions derived from validating the proposed concepts.
Through this empirical work, we expand the design space of interactive dance practice systems from the established context of single-user practice (primarily focused on gesture recognition) to a multi-user, group-based scenario focused on feedback and communication.

\end{abstract}

\ccsdesc[500]{Human-centered computing~Scenario-based design}
\ccsdesc[500]{Human-centered computing~Participatory design}
\ccsdesc[500]{Human-centered computing~Empirical studies in interaction design}
\keywords{Offline group dance practice; Cheerleading; Group-level performance; Pose estimation; Group communication; Learning}


\begin{teaserfigure}
  \includegraphics[width=\textwidth]{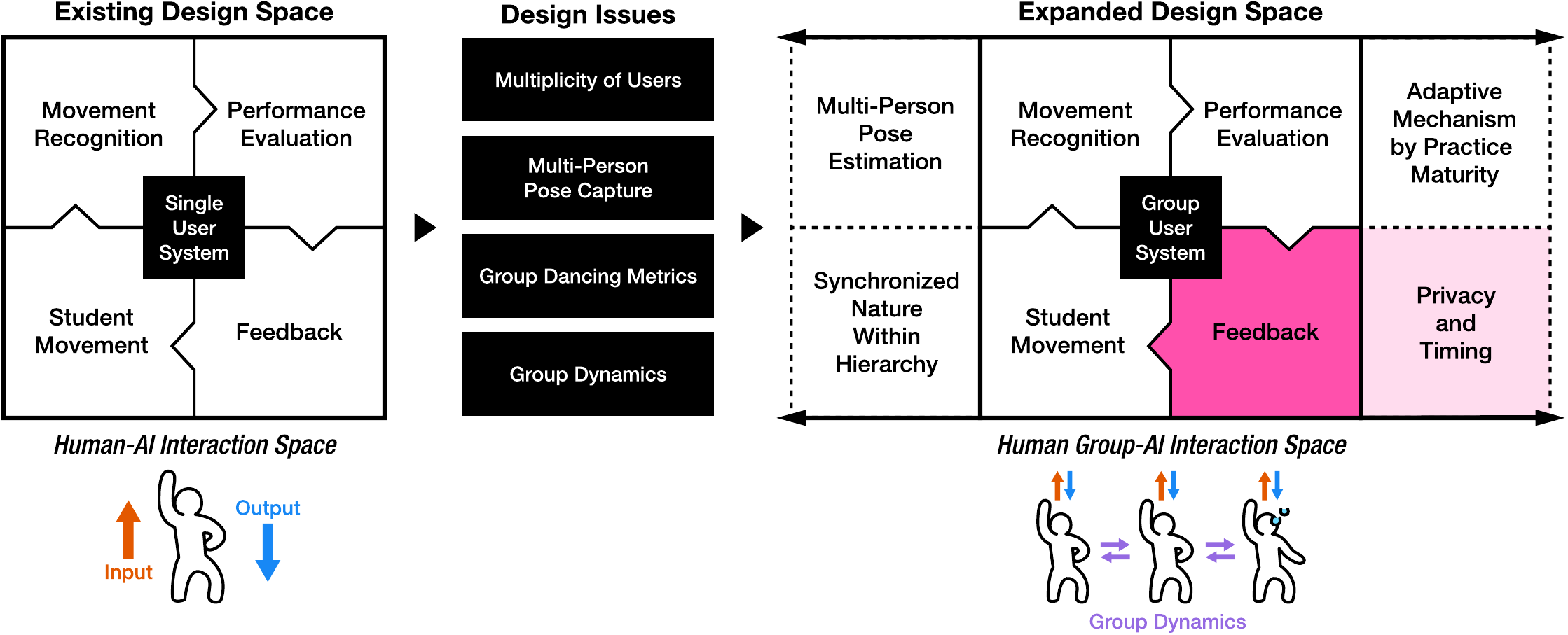}
  \caption{We explored and proposed new dimensions for the design of a group-based interactive dance practice system. The group contexts require consideration of different feedback strategies such as timing and privacy.}
  \Description{This diagram contrasts two design spaces for dance technology. The left side, "Existing Design Space," focuses on individual user interaction with components like Movement Recognition, Performance Evaluation, and Feedback. The right side, "Expanded Design Space," extends to a group system, incorporating Multi-Person Pose Estimation and Group Dynamics. It emphasizes advanced features such as adaptive mechanisms by practice maturity, and considerations for privacy and timing, reflecting the shift from solo to group dance activities in a Human Group-AI Interaction Space.}
  \label{fig:teaser}
\end{teaserfigure}


\maketitle

\input{sections/01_Introduction}

\input{sections/02_RelatedWork}

\input{sections/04_ExistingDesignSpace}
\input{sections/05_FormativeStudy}

\input{sections/06_ExpandDesignSpace}
\input{sections/07_EvaluateDesignSpace}
\input{sections/08_Discussion}
\input{sections/09_Limitations_FutureWork}
\input{sections/10_Conclusion}

\begin{acks}
The authors deeply appreciate the UNICH, UNIST Cheerleading Crew, for participating in this study and Eunyong and Yonghwan for valuable discussions. This research was partially supported by a grant from the Korea Institute for Advancement of Technology (KIAT) funded by the Government of Korea (MOTIE) (P0025495, Establishment of Infrastructure for Integrated Utilization of Design Industry Data). This work was also partially supported by the Technology Innovation Program (20015056, Commercialization design and development of Intelligent Product-Service System for personalized full silver life cycle care) funded By the Ministry of Trade, Industry \& Energy(MOTIE, Korea).
\end{acks}

\bibliographystyle{ACM-Reference-Format}
\bibliography{sample-base}

\end{document}

%% file: sections/01_Introduction.tex
\section{Introduction}
Group dancing involves teams of performers engaged in highly synchronized and tightly choreographed physical action. Such dances are a longstanding part of most human societies---they appear in many cultures and across most of recorded history~\cite{sachs1963world}. They persist and prosper today in a variety of currently popular forms, such as traditional folk dancing~\cite{chauvigne2019multi}, line dancing~\cite{Joseph19}, and cheerleading~\cite{Lesko88}. Cheerleading, a spirited amalgamation of synchronized dance, acrobatics, and chants, is a particularly relevant and topical example of the form due to its broad global appeal, the relatively young age of its practitioners, its emphasis on encouraging female participation in athletics~\cite{Moritz11} and its recent provisional recognition as an Olympic sport~\cite{Ruiz16}. The benefits of participating in amateur cheerleading and other forms of group dance are also substantial and extend well beyond improved physical fitness~\cite{Thomas2004} to include greater self-esteem~\cite{Deng2022} and group bonding (in the form of elevated levels of pro-social behavior associated with performing synchronized movements~\cite{CIRELLI18}). 

Despite these benefits, mastering the intensive synchronized movements of cheerleading proves challenging for many participants~\cite{zhou_syncup_2021}. Such failures are associated with dropping out of training and, more seriously, effects such as reduced confidence, self-esteem, and perceived social standing~\cite{Barnett06}. We believe interactive solutions to facilitate practicing group dance routines could improve novices’ performance, prolong their engagement with group dance activities, and, ultimately, increase the rewards they reap from their practice. However, while the HCI literature tackling dance learning and education is substantial, covering topics as diverse as providing accurate and consistent feedback on performed movements~\cite{gibbons_feedback_2004}, through improving support for self-observation~\cite{raheb_hci_2016}, to exploring feedback forms and modalities~\cite{trajkova_takes_2018}, it predominantly considers dance learning to be a solo activity. It focuses on and seeks to support individual dancers practicing alone. We identify a research gap in the lack of literature considering the unique aspects, properties, and environment of group dance activities. These range from the fundamental and functional, such as the core emphasis on synchronized group movements~\cite{zhou_syncup_2021}, through the social, such as the delivery of appropriate and supportive collective feedback in a group setting~\cite{zhu_zoombatogether_2023}, to the instructional, in that many cheerleading troupes lack professional trainers and instead educational activities are simply conducted by the most experienced members~\cite{Kim2023Action}. 

This paper seeks to fill this gap and understand the learning needs of troupes engaged in group dance practice. Additionally, we seek to design interactive systems that can support real-world group dance training. In this work, we specifically focus on dance cheerleading, a sub-genre distinguished from other cheerleading forms by its strong focus on rhythmic, synchronized movements and reduced emphasis on individualistic expression~\cite{deng2022effects}. It is primarily practiced by college teams, in groups of between about 10 and 30 individuals, who lack professional trainers. These properties make it an ideal candidate for study – the precision of the movements it requires is high, the social environments in which it is practiced are complex and nuanced, and it lacks an established tradition of professional coaching. It represents a dance learning context that has been historically overlooked and one in which we believe that interactive technology has the potential to realize substantial gains in terms of learner performance, attitude, and experience.

To realize these goals, we first conducted a formative study using video observation of practice sessions and one-on-one interviews with a cohort of 16 amateur cheerleaders (five instructors and 11 learners). We synthesized the data captured into 24 distinct challenges under the overarching theme of giving and receiving feedback on learners’ performance. We took these challenges forward into a design workshop conducted with six design researchers, who ultimately generated 55 ideas for vision-based interactive systems capable of meeting one or more of the identified challenges. To validate these ideas, we weaned them down and transformed them into 5 technology probes and 15 representative storyboards, each encompassing aspects of several of the original ideas, and presented them to nine cheerleaders (organized in groups of three) following the technology probe~\cite{hutchinson_technology_2003} and speed dating~\cite{davidoff2007rapidly, zimmerman_speed_2017} approach. Their feedback provides insight into both the specific designs presented and the underlying issues they seek to address and represents valuable guidance for future work seeking to develop interactive systems to support group dance activity. 


\begin{figure*}[]
  \centering
  \includegraphics[width=1.0\textwidth]{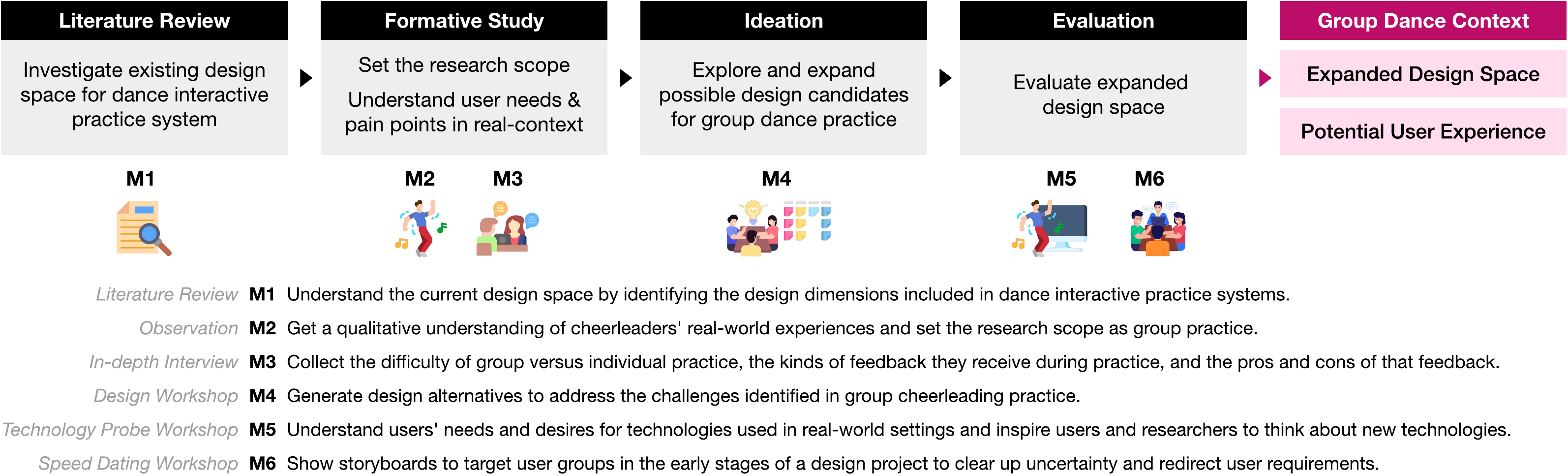}
  \caption{Schematic of the research process: The purple box shows the process we followed to create concrete prototypes, from understanding end users and their context to exploring and evaluating the pre-prototype design space. Finally, we suggest future directions for system design and development.}
  \Description{This displays a research process diagram for developing group dance interactive practice systems. It details steps from literature review to potential user experience enhancement, using method icons (M1 to M6) such as literature review, observations, interviews, design workshop, and technology probe & speed dating workshops, culminating in evaluating expanded design concepts.}
  \label{fig:researchOverview}
\end{figure*}

This study makes the following contributions to the DIS community:
\begin{itemize}
\item We expand the design space of interactive systems to support teaching and learning dance by documenting the unique challenges inherent in group contexts (Section 5). Future researchers can use this content to understand the practice of group dance learning and develop their own novel prototypes and designs. 
\item We present novel designs for supporting group dance learning (Sections 6 and 7), providing a concrete case study of how such systems can be conceived and developed. Our solutions focus on feedback and communication and are validated by developing 15 specific scenarios and using these in speed dating and technology probe studies. These design examples and user reactions can catalyze and inspire future designers working on interactive systems for learning group dance. 
\end{itemize}
This paper is structured as follows (\autoref{fig:researchOverview}). Firstly, we applied a human-centered approach to deeply understand human-human interactions (HHI) \cite{pritchard1988effects, weldon1993group}. We show the relevance and novelty of our proposal in addition to the existing design space through the literature review (\autoref{fig:researchOverview}-M1) (Section 3) on this topic. We then conducted a formative study that resulted in addressing 24 specific needs and pain points (\autoref{fig:researchOverview}-M2\&M3) (Section 4). We then conducted a design ideation workshop with six designers, which resulted in 55 innovative ideas that we refined into 15 storyboards to articulate new design dimensions (\autoref{fig:researchOverview}-M4) (Section 5). For the evaluation phase, we tested these dimensions against potential user experiences using technology probes and speed-dating workshops focused on RGB camera-based designs (\autoref{fig:researchOverview}-M5\&M6) (Section 6). This approach ensured that our expansion of the design space was theoretically grounded and deeply rooted in the empirical evidence from real-world group dance practice sessions. Finally, we discuss the design implications (Section 7) and conclude the paper (Section 8).

%% file: sections/02_RelatedWork.tex
\section{Related Work}
\subsection{Dance Practice and Learning in HCI}
Dance education, rich in diversity across various genres, presents a unique set of challenges in both individual and group contexts. While individual dance practice emphasizes mastering specific techniques, personal expression, and interpreting music, it inherently involves navigating physical and psychological challenges such as maintaining endurance and coping with the pressures of solo performance \cite{de2023psychological, etzinger2023dance, schwender2018effects}. However, these challenges take on a different dimension in the context of group dance. Specifically, despite the critical role of group dynamics in dance \cite{Buckroyd2000group}, the particular difficulties encountered in group settings — such as achieving synchronization, managing interdependence among dancers, and navigating group feedback processes — have been, by and large, overlooked in the existing literature on interactive systems to support dance education.

The spectrum of group dance genres, from the coordinated precision of ballet ensembles to the collective storytelling in folk dances, highlights the importance of synchrony and teamwork. Achieving seamless integration of movements in these styles is a significant challenge, requiring individual skill and a deep understanding of group behavior and interaction \cite{Buckroyd2000group}. This study, therefore, zeroes in on the element of synchronization as a pivotal aspect of group dance, recognizing its paramount importance in the aesthetic and technical success of group performances.

Dance cheerleading, aligning closely with synchronized dance, offers a unique perspective within the group dance milieu. This style is distinguished from other cheerleading forms by its focus on rhythmic, synchronized movements and less emphasis on individualistic expression \cite{deng2022effects}. Especially the standardization and precision required in Korean college cheerleading routines make it a compelling subject for studying group dance dynamics \cite{Kim2023Action}. It represents an ideal case to explore the complexities and challenges of group synchronization and collective performance in dance. 
This exploration aims to shed light on the less-explored aspects of group dance practice, specifically addressing the difficulties in achieving unified group execution and the role of feedback in such settings.

\subsection{Systems for Interactive Dance Practice}
Interactive systems for dance practice, developed within the realm of HCI, have historically employed various technologies to capture and analyze movement \cite{raheb_dance_2019}. These include vision-based motion capture~\cite{chan_virtual_2011, usui_learning_2015, tsampounaris2016exploring, aristidou2014motion, aristidou2015folk}, depth cameras (such as Microsoft Kinect)~\cite{trajkova_takes_2018, anderson_youmove_2013, kyan_approach_2015, camurri2016system, alexiadis_evaluating_2011, marquardt2012super}, and inertial motion sensors~\cite{kitsikidis2014multi}. Each of these tools offers unique capabilities for detecting and interpreting dancers' movements, contributing to the rich tapestry of technological solutions available for dance training. Alongside capturing movements, a significant body of research has also focused on translating the results of movement assessment into appropriate and effective feedback \cite{trajkova_takes_2018, trajkova_e-ballet_2016}. This feedback is delivered through various modalities and different styles. It is widely acknowledged that carefully designed and appropriate feedback must be delivered to enhance a dancer's learning experience and the accuracy of their performance \cite{trajkova_takes_2018}.

However, transitioning these systems to support group dance practice presents distinct challenges, not least in multi-person movement capture and analysis. Recent developments in computer vision and deep learning have begun to address these challenges 
and multi-person pose estimation using RGB cameras \cite{bazarevsky_blazepose_2020, lugaresi_mediapipe_2019, xu_monoperfcap_2018, zhang_pose-guided_2019}, which are now common in mobile phones, is now achievable. While previous approaches to capturing multiple people's movements through a variety of devices have been proposed, utilizing only a single RGB camera is an important step toward creating systems that are accessible and portable enough for real-world group dance environments \cite{lee_pose_2020, zhou_syncup_2021, zhu_zoombatogether_2023, kang_dancing_2023}. Accordingly, we adopt this approach throughout our work. Despite these technical developments, interactive systems to support group dance~\cite{zhou_syncup_2021, zhou_visualizing_2019, lee_cheerup_2023, zhu_zoombatogether_2023} remain scarce in the research literature in comparison to the wealth of work focusing on individual dance practice. This research gap, especially in creating accessible systems for group dance practice, forms the basis for the work described in this paper. 

\subsection{Early-Stage Design Evaluation}

Evaluating design alternatives at an early stage 
serves to explore and evaluate the extended design space before prototyping, ensuring alignment with user needs and expectations. This process employs methodologies such as speed dating workshops~\cite{zimmerman_speed_2017, davidoff2007rapidly}, experience prototyping~\cite{davidoff2007rapidly}, paper prototyping~\cite{sefelin2003paper}, and scenarios~\cite{interaction2016scenario}. It de-risks design ideas and thus serves as a crucial step in the design process. It enables designers to refine and adjust concepts based on practical user requirements, significantly enhancing the potential for the solutions to meet real-world needs effectively and without necessarily being tied to introducing new technologies into unexplored areas.

This study leverages a technology probe workshop~\cite{hutchinson_technology_2003, zhou_dance_2021} and the speed dating method~\cite{zimmerman_speed_2017, davidoff2007rapidly} to evaluate the use of RGB cameras in group dance contexts. Analogous approaches have been used in previous research on mixed reality mirror systems for improvisational choreography to quickly identify limitations and opportunities in the user experience of the technology~\cite{zhou_here_2023}. By enabling direct interaction with emerging technologies, the technology probe workshop helps identify user needs and the practical challenges of technology integration. Conversely, the speed dating methodology facilitates quick, iterative feedback on various design concepts, allowing for an agile refinement process based on user interaction and response. Together, they provide a robust framework for assessing both the technical viability and user experience implications of integrating RGB cameras into dance practice, guiding our design process to be both technologically innovative and deeply aligned with user needs. This dual approach ensures our exploration into the expanded design space is comprehensive and that the resulting interactive systems are both technologically functional and highly relevant to the specific demands of group dance practice.

%% file: sections/04_ExistingDesignSpace.tex
\begin{figure*}[]
  \centering
  \includegraphics[width=1.0\textwidth]{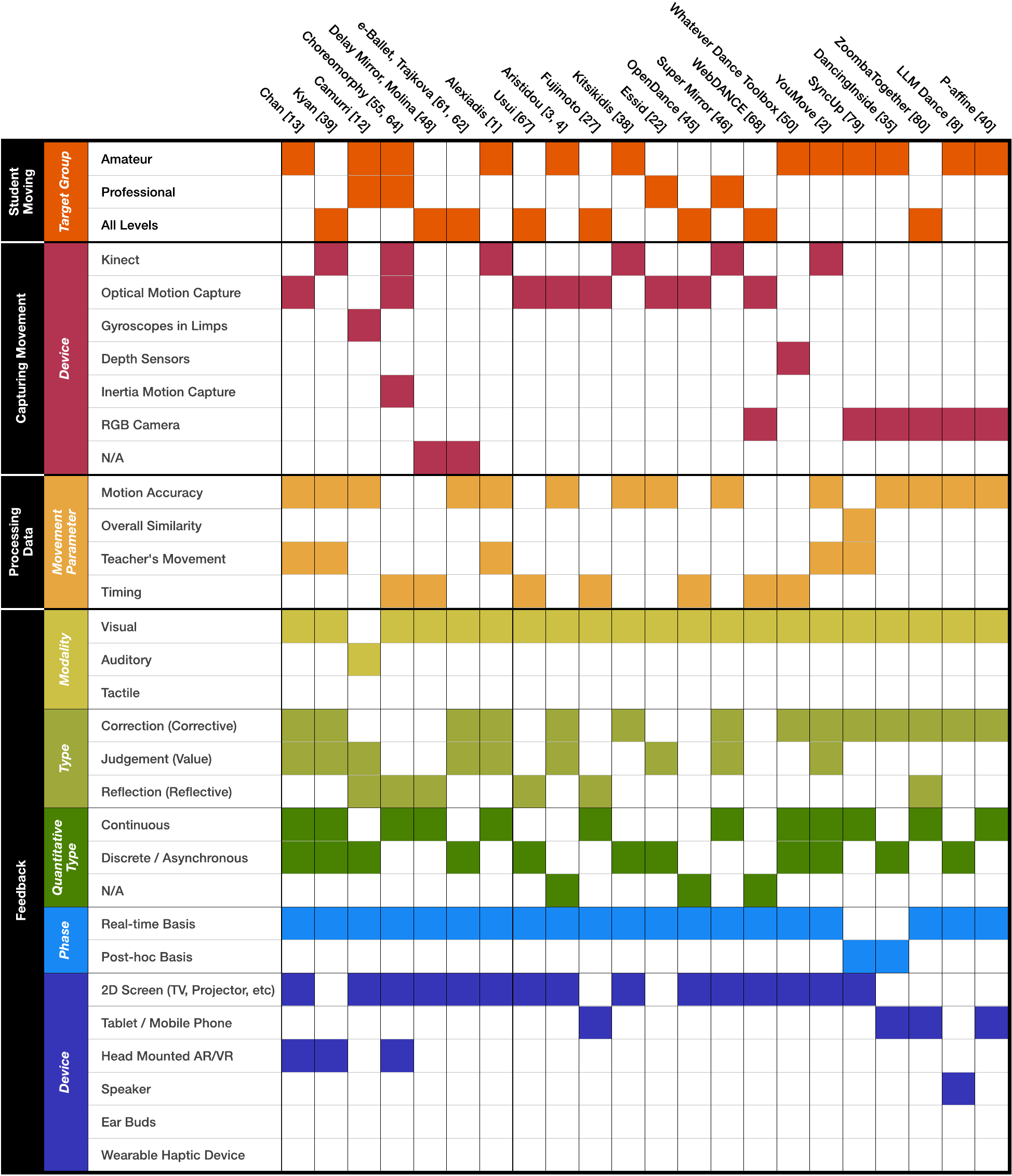}
  \caption{Existing Design Space: Overview of eight key design dimensions for interactive dance systems, based on a review of 22 studies. This diagram encapsulates essential factors like Device (Capture Movement), Movement Parameter, Modality, and others, highlighting the foundational elements for advancing dance technology in education.}
  \Description{This presents a comprehensive matrix categorizing the design space of interactive dance systems. The matrix is organized by various design dimensions such as devices used to capture movement (e.g., Kinect, optical motion capture), types of movement parameters analyzed (e.g., motion accuracy, timing), feedback modalities (e.g., visual, auditory, tactile), types of feedback (e.g., corrective, reflective), feedback cadence (continuous or discrete), and the devices for feedback delivery (e.g., 2D screen, wearable haptic device). Each cell in the matrix is color-coded to indicate the relevance or application of each element to different user groups (amateur, professional, all levels) and various technology aspects in dance education.}
  \label{fig:existingDesignSpace}
\end{figure*}

\section{Existing Design Space for Interactive Dance Systems}
We examined the existing design space for dance interactive practice systems by analyzing a total of 22 interactive concepts, 
including 17 concepts \cite{chan_virtual_2011, kyan_approach_2015, camurri_system_2016, raheb2018choreomorphy, molina2017delay, trajkova_takes_2018, alexiadis_evaluating_2011, usui2015learning, aristidou2014motion, aristidou2015folk, fujimoto2012dance, kitsikidis2014multi, essid2013multi, magnenat2008learning, marquardt2012super, bakogianni2007teaching, karkou2008traditional, pristavs2016badco, anderson_youmove_2013} from a 'Dance Interactive Learning System' review up to 2019 \cite{raheb_dance_2019} and
an additional five concepts introduced between  2019 and 2023 \cite{zhou_syncup_2021, kang_dancing_2023, zhu_zoombatogether_2023, blanchet2023integrating, lee_pose_2020}. This analysis led us to summarize the design dimensions of current systems into eight distinct dimensions, categorized within four broader categories reflecting the 'Dance Interactive System Workflow~\cite{raheb_dance_2019}': \textit{Student Moving, Capturing Movement, Processing Data and, Feedback}. These are shown in \autoref{fig:existingDesignSpace}, and the design dimensions are defined as follows:

\begin{itemize}
    \item \textit{Target Group}: Systems are designed to cater to various skill levels, including amateur dancers seeking foundational skills, professional dancers focusing on advanced techniques, and systems accommodating all levels, offering a broad range of functionalities. 
    \item \textit{Device(Capturing Movement)}: A variety of devices are employed for movement capture, including Kinect for gesture and posture recognition, optical and inertia motion capture systems for precise movement tracking, gyroscopes embedded in limbs for orientation and rotation data, depth sensors for spatial analysis, and RGB Cameras for accessible, camera-based capture. 
    \item \textit{Movement Parameter}: Analyzed parameters include motion accuracy (precision of movements), overall similarity (comparative analysis with a standard), teacher’s movement (mimicry and learning), and timing (temporal alignment with music or choreography). 
    \item \textit{Modality}: Feedback is delivered through various modalities, such as screen displays, auditory cues via speakers or earbuds, and tactile sensations through wearable haptic devices, catering to different learning preferences. 
    \item \textit{Type}: The feedback type encompasses correction (corrective feedback on errors), judgment (evaluative comments or scores), and reflection (encouraging self-assessment and insight). 
    \item \textit{Quantitative Type}: Feedback can be continuous, offering ongoing data during the dance, or discrete/asynchronous, providing information at specific intervals or after the performance. 
    \item \textit{Phase}: The timing of feedback varies, with some systems offering insights on a real-time basis during the dance and others on a post-hoc basis, analyzing performance after completion. 
    \item \textit{Device(Feedback)}: Devices for feedback delivery include 2D screens (TVs, projectors) for visual feedback, tablets/mobile phones for portable access, head-mounted AR/VR for immersive experiences, speakers for audio cues, earbuds for private auditory feedback, and wearable haptic devices for physical sensations.
\end{itemize}

Each of these dimensions plays a crucial role in defining the functionality and user experience of dance interactive systems functionality and user experience. Understanding these dimensions allows us to grasp the current landscape of such systems, providing a foundation for our study's objective to expand this design space for group dance contexts.


%% file: sections/05_FormativeStudy.tex
\section{Understanding the Problem Space of Group Cheerleading Practice}
Through an exploratory study, we aimed to uncover the challenges college student cheerleaders face during their practice sessions. Due to ease of access, we focused on the activities of a single cheerleading crew based at our local university. We engaged this group using complementary methods: indirect and participatory observation of practice sessions and in-depth interviews. The observational methods supported gathering insights from actual practice sessions, capturing cheerleaders' authentic and natural behavior while avoiding the recall biases that might impact self-reported data. The in-depth interviews delved into the participants' experiences, challenges, and perspectives, providing deeper insights into their pain points and shedding light on the behaviors seen during the observation sessions.

\begin{figure}[]
  \centering
  \includegraphics[width=0.4\textwidth]{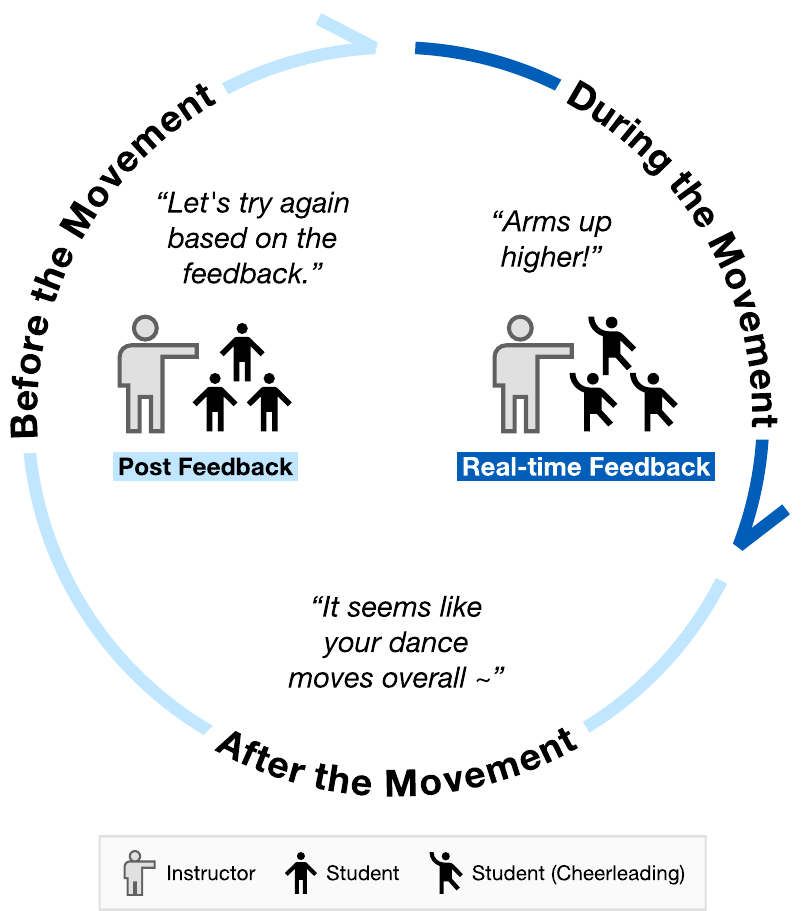}
  \caption{Micro-perspective View of Practice: The instructor gives \raisebox{-0.4ex}{\includegraphics[height=1.0em]{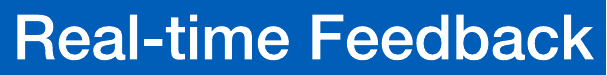}} during the small group practice. At the end of the practice, \raisebox{-0.4ex}{\includegraphics[height=1.0em]{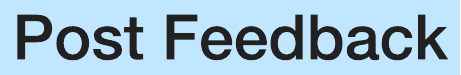}} is given on the practice, which is used to prepare for the next practice. So, in a real-world small group practice, post feedback includes feedback after the choreography practice and before the next practice.}
  \Description{This depicts a micro-perspective view of a dance practice cycle focusing on the timing and type of feedback an instructor provides. It illustrates three feedback phases during dance practice: Before the Movement-- The instructor provides post feedback, analyzing the previous attempt and suggesting improvements. During the Movement-- Real-time feedback is given, with immediate corrections such as "Arms up higher!" After the Movement-- A general review of the practice session is discussed to prepare for the next practice. The figure shows that from a micro perspective, there are three phases: before, during, and after the movement, but there is no real distinction between feedback before and after the movement. Therefore, it emphasizes that feedback can be divided into two types: real-time feedback and post-feedback.
}
  \label{fig:microperspectivePractice}
\end{figure}

\subsection{Indirect Observation}
We observed nine practice sessions with the cheerleading crew spread over three weeks. Each practice session lasted approximately two hours and was attended by 
16 individuals. Eleven of the troop members were first-year students (and cheerleading novices), while the remaining five were sophomores who had at least one year of cheerleading experience. This crew was organized according to experience level and grade, with one of the sophomores serving as president and two as vice presidents. In each session, we observed, one author acted as an observer. Additionally, we installed three GoPro cameras to support later analysis (based on indirect observation). Two were placed in the corners of the practice room to provide a wide-angle view of the entire space and encompass the movements of all participants. The final camera was attached to the chest of one of the sophomore peer instructors to provide a first-person view of their perspective.







\subsection{In-depth Interview Procedure}


Participants for the in-depth interviews were selected from students who are or have been involved in the cheerleading crew, including the 11 current student members (5M/6F, p1\textasciitilde p11) and 5 participants (3M/2F, p12\textasciitilde p16) who served in instructor roles (two of them retired). The students had an average age of 18.64 (SD=0.81), with an average of 4.64 months (SD=3.98) of cheerleading experience. Two students were sophomores with over a year of experience, while the other nine were freshmen with less than six months of experience. The instructors had an average age of 20 (SD=1.73), with 13.2 months (SD=7.46) of experience as a student and 6.4 months (SD=4.16) as an instructor in cheerleading. Both students and instructors had experience in other group dances such as ballet or K-pop, with an average of 9.82 and 14.8 months, respectively. All participants, except the two retired instructors, participated in the observation phase. This multidimensional selection aimed to capture insights from both the instructors' and students' perspectives.

Interviews were conducted individually with each participant in the troop, ensuring perspectives from senior and junior members were captured. Sessions lasted approximately one hour. Interviews were audio recorded and transcribed with participant consent. In general, the interviews focused first on seeking explanations for behaviors seen in the participant observation sessions and second on how participants anticipated that interactive systems could be designed to support cheerleading activities in the future. 



\begin{figure*}[]
  \centering
  \includegraphics[width=1.0\textwidth]{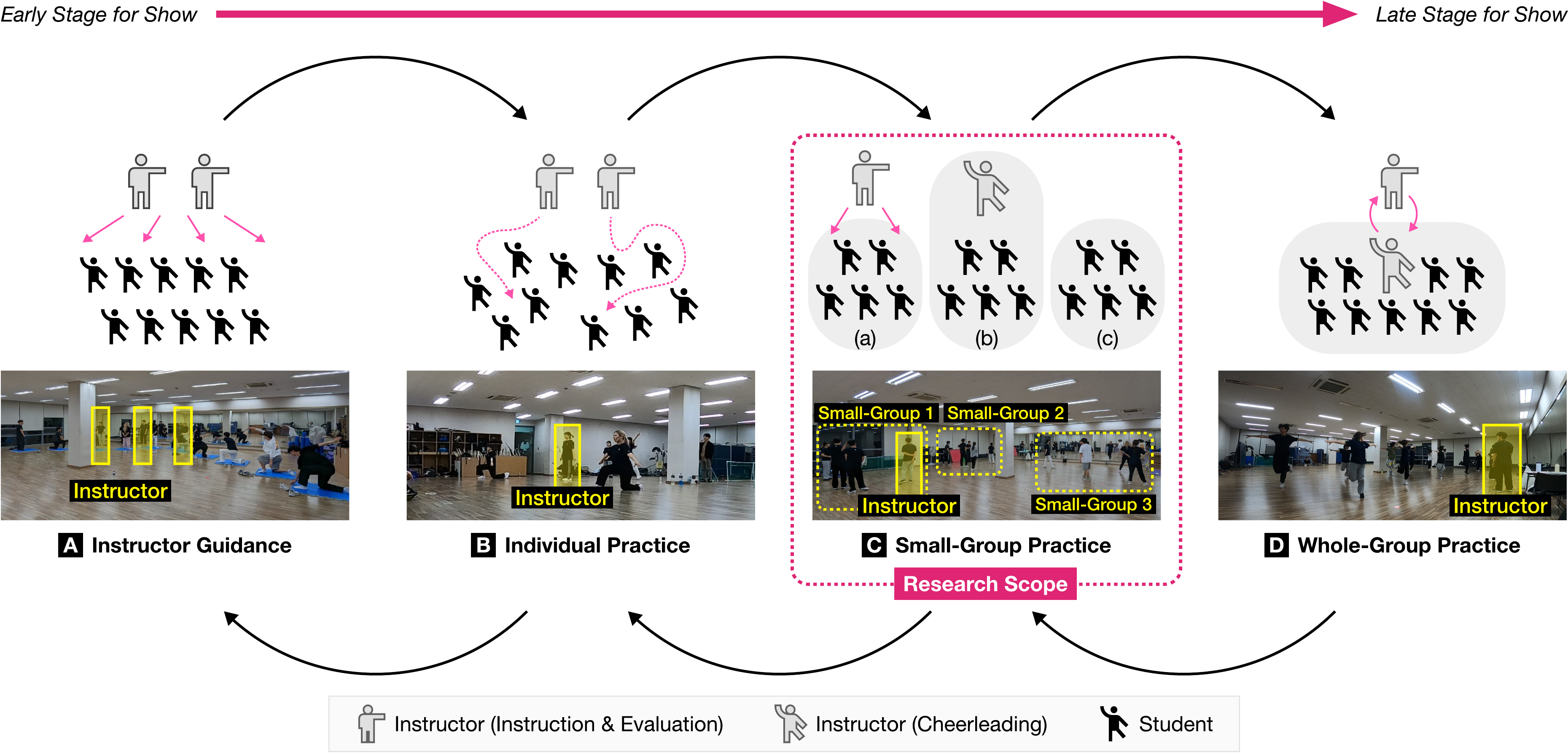}
  \caption{Macro-perspective View of Practice: If each step goes well, the dancers move on to the next step. If each step is unsuccessful, they return to the previous step. \raisebox{-0.4ex}{\includegraphics[height=1.0em]{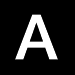}} Initially, the instructors teach the dancers the choreography. \raisebox{-0.4ex}{\includegraphics[height=1.0em]{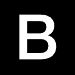}} Based on this, the dancers learn the choreography on their own. At this time, the instructors may go around and give individual coaching. \raisebox{-0.4ex}{\includegraphics[height=1.0em]{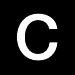}} Once the movements have been learned, the dancers practice in groups of 3 to 5. Small-group practice can be (a) under the instructor's guidance, (b) with the instructor dancing together, or (c) just the dancers without the instructor. \raisebox{-0.4ex}{\includegraphics[height=1.0em]{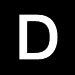}} In the final stage, everyone comes together to synchronize the entire movement. At this time, the instructor provides guidance or engages in exercises together.}
  \Description{It illustrates the stages of dance practice from early to late show preparation through a series of four panels: Instructor Guidance: An instructor demonstrates dance moves to a line of dancers directly facing him. Individual Practice: Dancers practice the moves independently without the instructor's intervention. Small-Group Practice: Dancers practice in small groups of three to five, shown in three sub-stages: with the instructor actively participating, the instructor observing but not participating, and without the instructor. Whole-Group Practice: All dancers practice together in unison, synchronizing movements under the instructor's guidance. Each panel is supported by a corresponding real-life image showing dancers and the instructor in a practice setting, emphasizing the transition from learning to execution.}
  \label{fig:macroperspectivePractice}
\end{figure*}

\subsection{Analysis of College Students' Cheerleading Practice}
We found that the troop frequently split into small practice groups (\autoref{fig:macroperspectivePractice}-\raisebox{-0.4ex}{\includegraphics[height=1.0em]{figures/C.pdf}}). Busy with sophomore-level practice, instructors often could not help junior members, especially during these small group sessions. In this manner, we found that students had difficulty knowing when and how to correct their behavior within the small groups. Interested in this issue, our interviews explicitly queried participants on the role of small group practice in the overall learning process. It quickly became apparent that while the end goal is for everyone to do the same thing, everyone cannot do the same thing at the start. Consequently, we found that the structure of overall macro-practice within the group iterated between periods of individual practice, small group practice, and complete practice, transitioning between these stages as and when required (\autoref{fig:macroperspectivePractice}).

We further investigated the characteristics and challenges of feedback in small group practice in the presence of an instructor. These insights emerged from thematic analysis~\cite{braun_using_2006, hsieh_three_2005} and fell into three broad categories: Insights into feedback received in real-time during the action, insights received after action, and common insights (\autoref{tab:insightsList}). These three categories reflect the observation that feedback in a non-professional dance studio takes place either during a movement or after a movement. While prior work has noted that practice can be categorized into three phases (before, during, and after a movement)~\cite{trajkova_designing_2019}, our experience in an amateur dance studio indicates that the line between 'before movement' and 'after movement' is highly blurred (Figure~\ref{fig:microperspectivePractice}). The final categorized insights were rendered onto cards and used in an ideation workshop in the next phase of the project (\autoref{fig:designWorkshop}).

\begin{table*}[]
  \caption{24 insight list from the in-depth interview: Summary of 24 insights from the formative study on feedback experiences in group dance practice. The table categorizes insights into three key phases: during movement, post-movement, and those applicable across both contexts. This classification highlights dancers' nuanced preferences and challenges in receiving feedback, providing a foundational understanding for designing interactive dance feedback systems.}
  \label{tab:insightsList}
  \centering
  \includegraphics[width=1.0\textwidth]{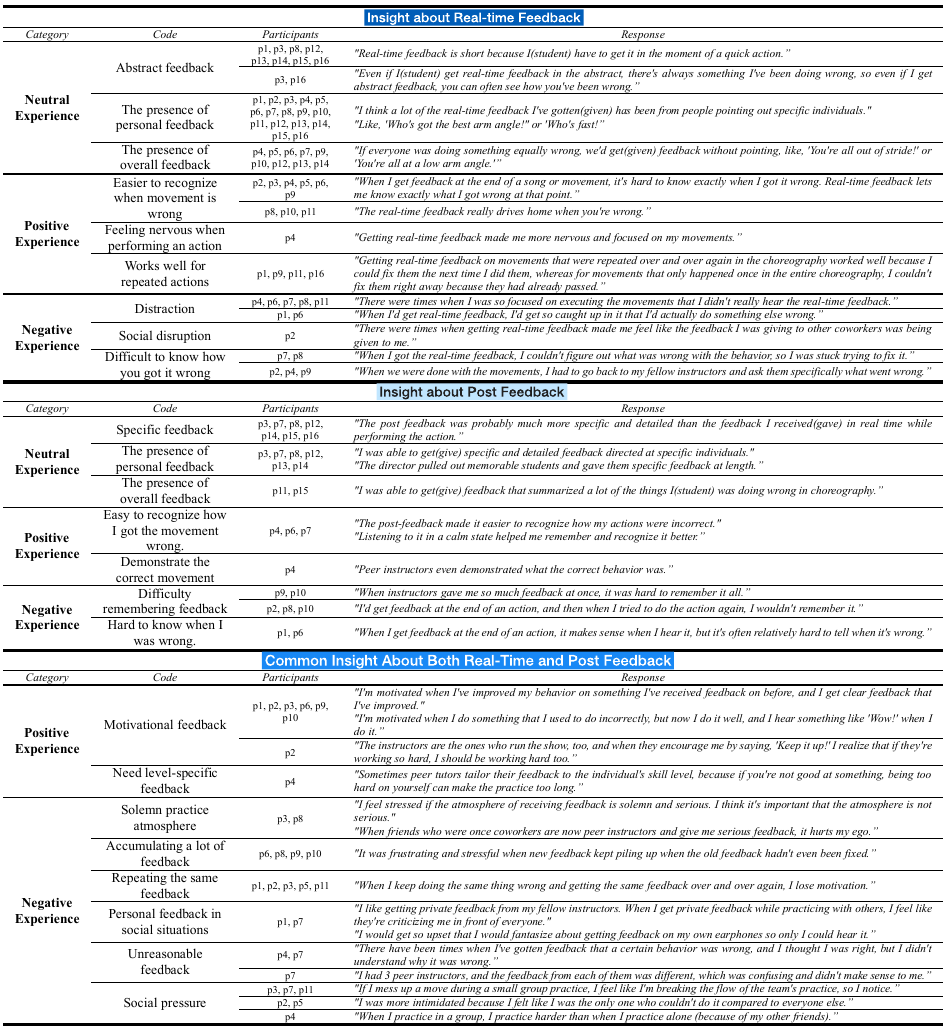}
  \Description{The table provides insights from an in-depth interview on feedback in dance practice, structured by feedback timing: real-time and post-movement. It categorizes experiences into positive, neutral, and negative experiences, detailing participants' insights and reactions to different feedback scenarios. The table aims to inform the design of interactive dance feedback systems by highlighting dancers' preferences and challenges.}
\end{table*}

%% file: sections/06_ExpandDesignSpace.tex
\section{Expanding the Design Space of Interactive Systems for Group Dance Practice.}
This section details the systematic process we used to generate prospective designs for interactive systems that can improve small-group cheerleading practice. 
We outline the workshops and methods we used that were instrumental in generating, diverging, and converging ideas. Our goal was to, firstly, expand the design space of interactive systems for dance practice into a set of provocative ideas tackling genuine problems and, secondly, to converge this set down into a smaller group of firm and clearly expressed concepts that would be amenable to further rounds of design inquiry, development, and study. 

In particular, we chose a computer vision-based approach to explore and expand the design space for improving small-group cheerleading practice. Specifically, we chose to analyze body movement through RGB single-camera-based pose estimation. The portability of the single-camera setup made it suitable for a real-world dance studio environment. We also provided participants with information that can be computed from pose coordinates (pose similarity\cite{zhou_syncup_2021, lee_cheerup_2023}, pose correctness\cite{kang_dancing_2023, lee_pose_2020}, temporal alignment\cite{zhou_syncup_2021}, etc.). 

\begin{figure*}[]
  \centering
  \includegraphics[width=1.0\textwidth]{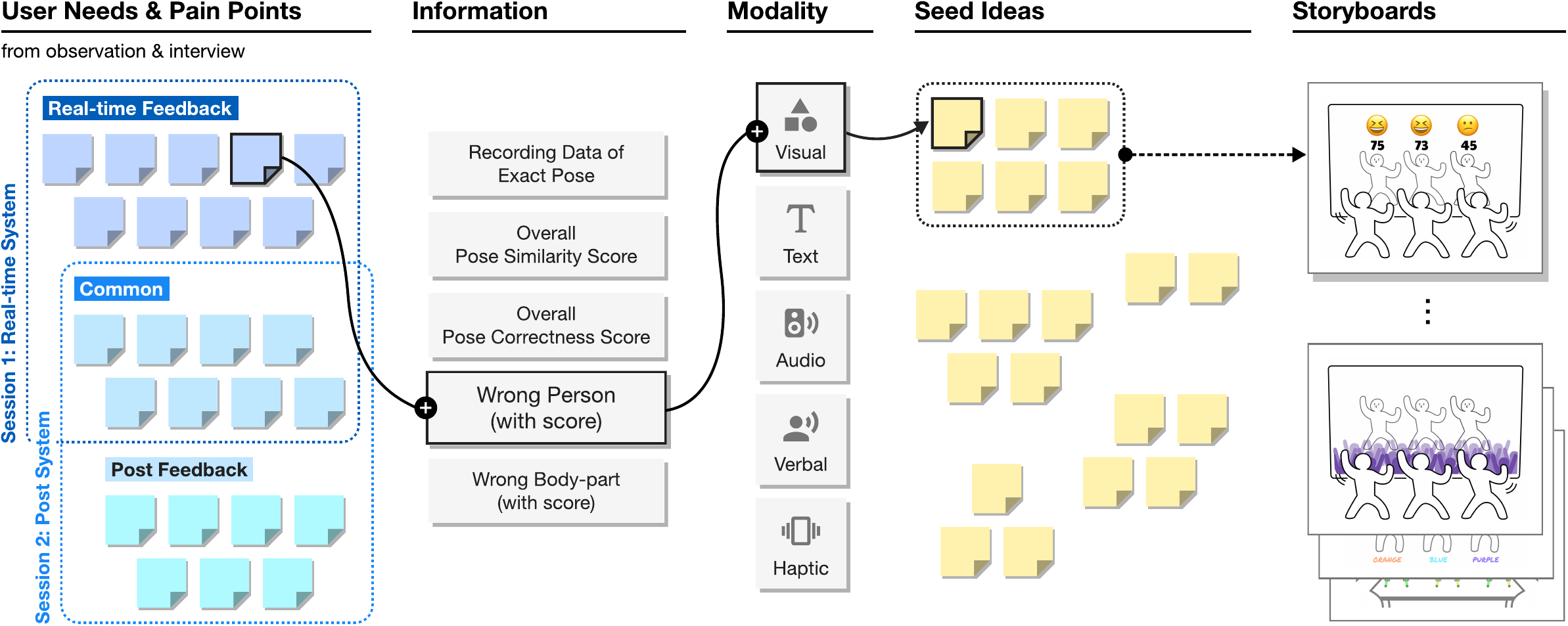}
  \caption{Workflow of the Idea Divergence Workshop Process: This diagram visualizes the development of a dance feedback system, beginning with user insights from interviews that inform critical feedback information. These insights guide the selection of feedback modalities, culminating in creating seed ideas grouped into final storyboards showcasing potential system interactions. In Session 1, ideation for \raisebox{-0.4ex}{\includegraphics[height=1.0em]{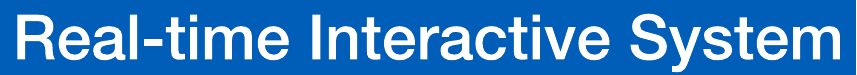}} was conducted based on \raisebox{-0.4ex}{\includegraphics[height=1.0em]{figures/realtime_feedback.pdf}} insights and \raisebox{-0.4ex}{\includegraphics[height=1.0em]{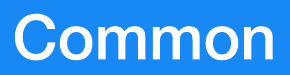}} insights. In Session 2, ideation for \raisebox{-0.4ex}{\includegraphics[height=1.0em]{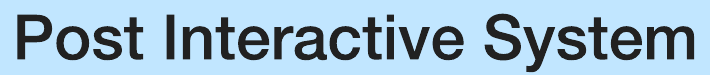}} was conducted based on both \raisebox{-0.4ex}{\includegraphics[height=1.0em]{figures/post_feedback.pdf}} insights and \raisebox{-0.4ex}{\includegraphics[height=1.0em]{figures/common.pdf}} insights.}
  \Description{This outlines the Idea Divergence Workshop Process for a dance feedback system, detailing two sessions: one for Real-time Feedback and another for Post Feedback. It starts with user needs and pain points identified through observation and interview, leading into specific information categories like pose correctness and similarity scores. These feed into modalities for delivering feedback—visual, text, audio, verbal, and haptic. The process culminates in seed ideas, illustrated by storyboards, which visualize potential user interactions and feedback presentation in the system.}
  \label{fig:designWorkshop}
\end{figure*}

\subsection{Idea Divergence Workshop}
This workshop 
focused on developing ideas for providing effective feedback to a small group of cheerleading practitioners that could address or highlight the experiences reflected in the existing 24 insight lists from the previous step. The main objective was to explore different ways the 'who, what, when, and how' could be communicated through an interactive system based on assessing the cheerleaders' behavior during small group practice (\autoref{fig:designWorkshop}). This ideation workshop involved six design graduate students(3M/3F) and had the following structure.

\textbf{Ice breaking \& priming (15 minutes)}:
The workshop began with a brief introduction to cheerleading and its unique challenges, especially in small group practice. The lack of clear feedback was highlighted as a problem, and examples of computer vision-based interactive systems for dancing were presented to set the stage for idea generation \cite{zhou_syncup_2021, lee_cheerup_2023, anderson_youmove_2013, zhou_here_2023, trajkova_takes_2018}.

\textbf{Introduce Insights \& Modality (5 minutes)}:
Participants were provided insight cards derived from the previous in-depth interviews: \raisebox{-0.4ex}{\includegraphics[height=1.0em]{figures/realtime_feedback.pdf}}, \raisebox{-0.4ex}{\includegraphics[height=1.0em]{figures/post_feedback.pdf}}, and \raisebox{-0.4ex}{\includegraphics[height=1.0em]{figures/common.pdf}} insights (Table \ref{tab:insightsList}). Participants were also introduced to different modalities (visual, auditory, tactile, etc.) that could be considered for delivering feedback.

\textbf{Brainstorming (45 minutes + 45 minutes)}:
The brainstorming session was divided into two parts: The first 45 minutes focused on systems that provide \raisebox{-0.4ex}{\includegraphics[height=1.0em]{figures/realtime_feedback.pdf}}; insight cards related to real-time and common insights were utilized. The second 45 minutes focused on systems that provide \raisebox{-0.4ex}{\includegraphics[height=1.0em]{figures/post_feedback.pdf}}, using cards related to post feedback and common insights. In both sessions, participants were asked to explore how and what information could be communicated to address or emphasize specific insights.

To expand the design space, we conducted an ideation workshop that generated 55 ideas for an interactive system to improve group cheerleading practice. 26 ideas focused on \raisebox{-0.4ex}{\includegraphics[height=1.0em]{figures/realtime_feedback.pdf}} during movement, and 29 focused on providing \raisebox{-0.4ex}{\includegraphics[height=1.0em]{figures/post_feedback.pdf}}. These ideas were systematically categorized based on timing, audience, information conveyed, devices used, and feedback modality. The raw data of the seed ideas is presented as supplementary material. We combined these ideas from this rich collection into a provocative storyboard that effectively represented the expanded design space (\autoref{fig:storyboards}). These storyboards served as a valuable tool for evaluating the potential and limitations of the proposed interactive system in the subsequent speed-dating workshop.

\subsection{Expanded Design Space}
We employed storyboard-based design representations to facilitate a nuanced understanding of potential user experiences and system interactions. Because of our focus on the small group context, these storyboards were curated based on three main criteria: 1) consideration of the group context, 2) anticipation of interactions between members after the system is deployed, and 3) inclusion of scenarios not adequately addressed in existing dance interactive learning systems.

Not only merely addressing the superficial challenges faced by cheerleaders, our selection of storyboards was intentionally provocative, aiming to catalyze discussion around the complex group dynamics we had observed. These storyboards also extended the themes and techniques explored in the Technology Probe Workshop, providing a cohesive narrative for vision-based interactive systems.

\begin{figure*}[]
  \centering
  \includegraphics[width=1.0\textwidth]{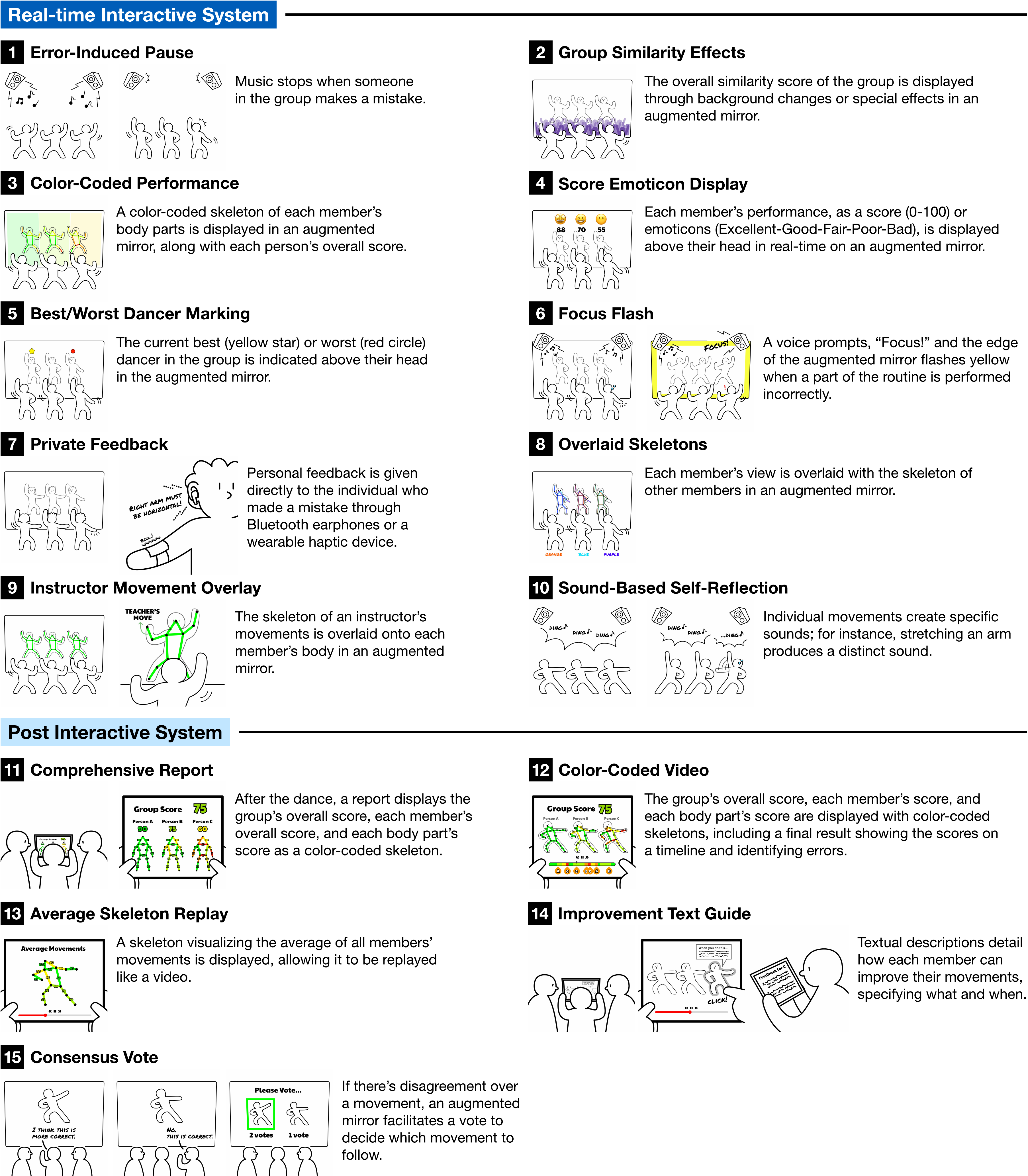}
  \caption{Storyboards: A compilation of 15 storyboard images depicting real-time and post-feedback concepts for interactive dance systems. The collection features ten ideas designed for \raisebox{-0.4ex}{\includegraphics[height=1.0em]{figures/realtime_interactive_system.pdf}} and five ideas for \raisebox{-0.4ex}{\includegraphics[height=1.0em]{figures/post_interactive_system.pdf}}, illustrating a broad spectrum of approaches for supporting group dance practice through vision-based technology.}
  \Description{This displays 15 storyboards split into real-time interactive and post-interactive systems for dance training. It visually explores advanced feedback mechanisms using augmented mirrors and other technologies. Examples include an Error-Induced Pause where music stops for mistakes, a Color-Coded Performance for accuracy visualization, and a Comprehensive Report that summarizes performance post-practice. These storyboards demonstrate potential feedback methods designed to enhance learning and precision in dance training through immediate and reflective feedback mechanisms.}
  \label{fig:storyboards}
\end{figure*}

\begin{figure*}[]
  \centering
  \includegraphics[width=0.80\textwidth]{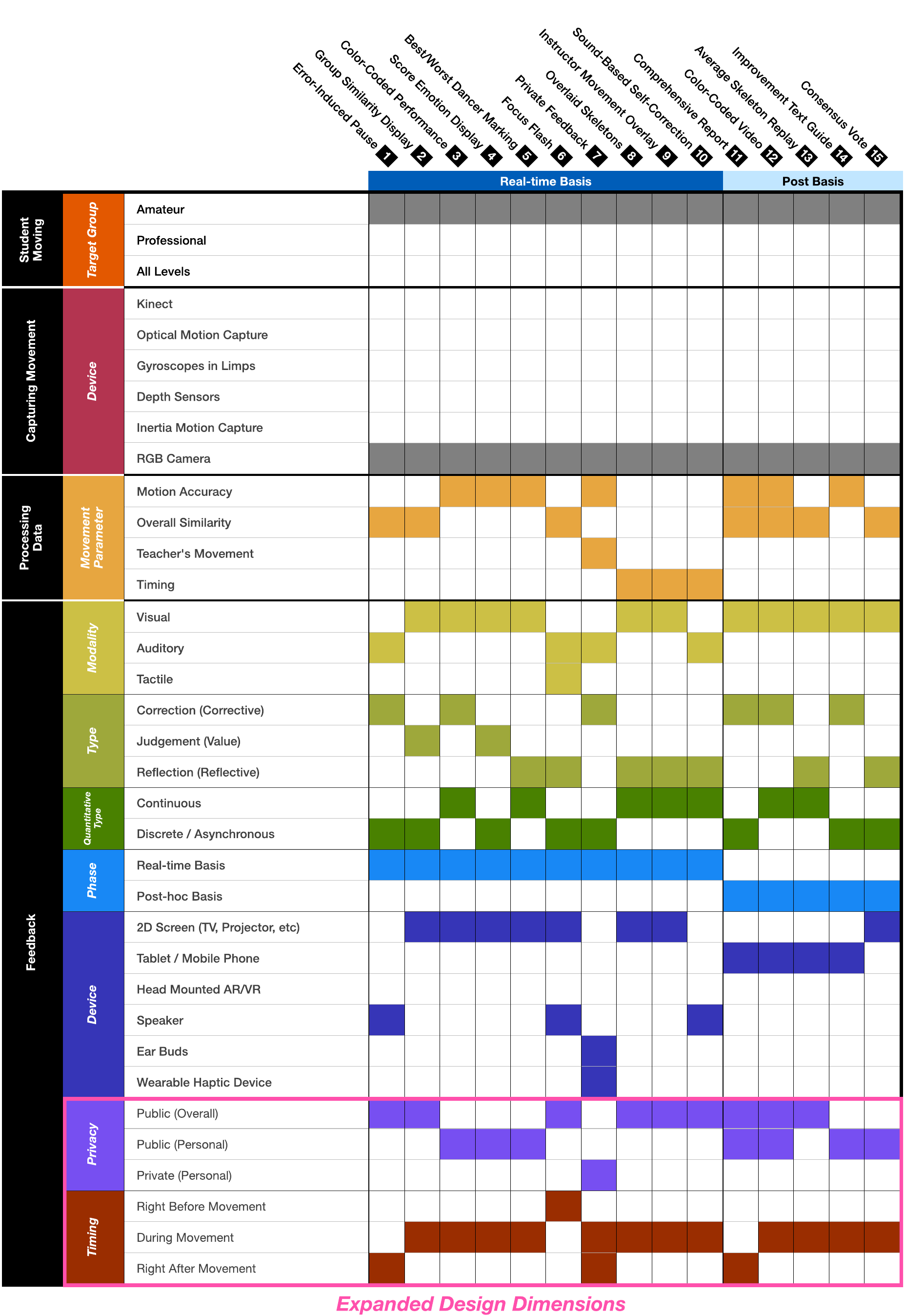}
  \caption{Expanded Design Space: The design dimensions for interactive dance systems are illustrated through 15 storyboards developed from a design ideation workshop. This expanded framework introduces two additional dimensions—Feedback Timing and Feedback Privacy—to the existing eight, showcasing the evolution and diversification of design considerations for enhancing group dance practice.}
  \Description{This matrix visualizes the design dimensions from 15 storyboards created in an ideation divergence workshop, highlighting newly added dimensions of feedback timing and privacy. It builds on the existing design dimensions shown in Figure 3, presenting a refined framework for interactive dance systems across various user levels and technology applications.}
  \label{fig:storyboardDimension}
\end{figure*}

To compare different interactive paradigms in a group cheerleading context, we crafted 15 storyboards. Of these, ten focused on providing \raisebox{-0.4ex}{\includegraphics[height=1.0em]{figures/realtime_feedback.pdf}} during movements, while five were geared towards delivering insights of \raisebox{-0.4ex}{\includegraphics[height=1.0em]{figures/post_feedback.pdf}} (\autoref{fig:storyboards}). A description of each storyboard and the types and modalities of feedback the system provides can be found in \autoref{fig:storyboardDimension}. 

%% file: sections/07_EvaluateDesignSpace.tex
\section{Evaluating Expanded Design Space}
We assessed the design space for interactive feedback systems to bolster group dance practice. This evaluation sought to pinpoint the limitations and opportunities of various interactive concepts and technologies generated from the preceding design ideation workshop. Our objective was to gather insights critical for guiding the development of tangible prototypes. Specifically, we concentrated on understanding how these systems could be effectively utilized within a real-world dance studio, exploring their potential applications and impact on the dance practice environment.

\begin{figure*}[]
  \centering
  \includegraphics[width=1.0\textwidth]{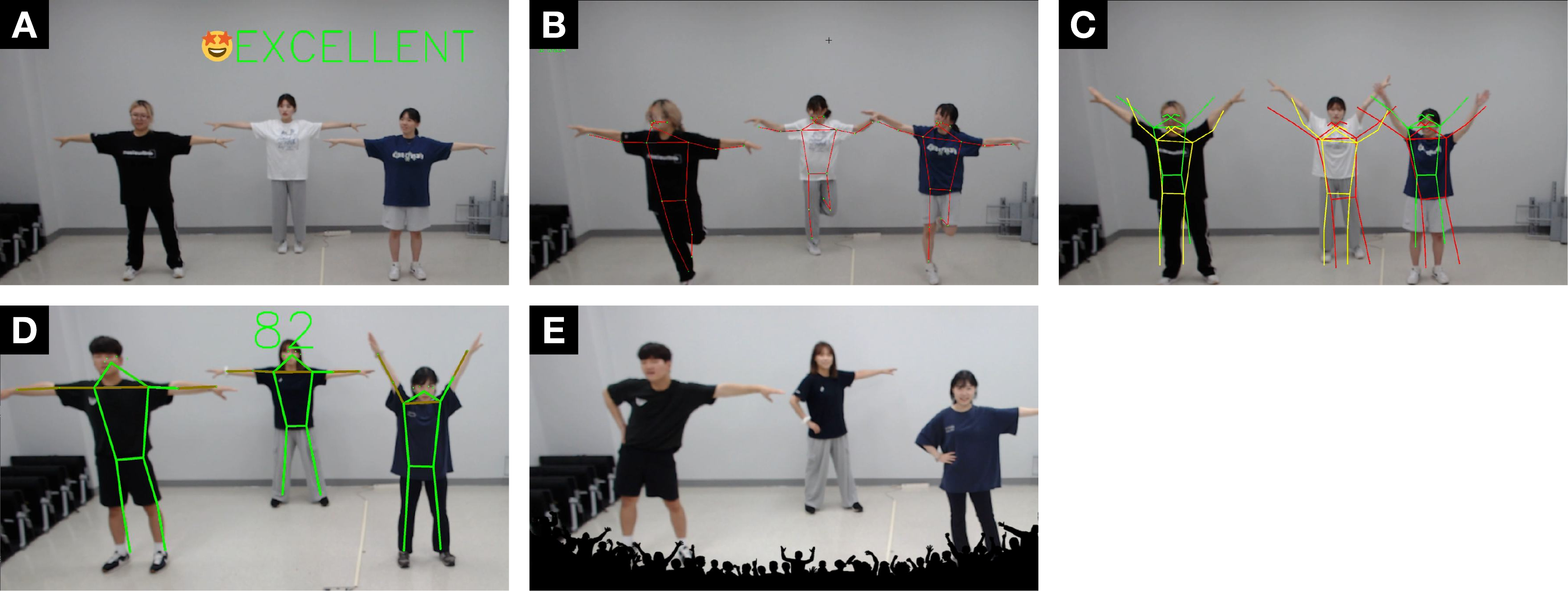}
  \caption{Technology Probes: Participants experienced the technology probe through an augmented mirror during the workshop. \raisebox{-0.4ex}{\includegraphics[height=1.0em]{figures/A.pdf}} Presenting overall pose similarity through emoticons, \raisebox{-0.4ex}{\includegraphics[height=1.0em]{figures/B.pdf}} overlaying pose skeletons on top of each participant, \raisebox{-0.4ex}{\includegraphics[height=1.0em]{figures/C.pdf}} overlaying each other's pose skeletons (not their own) on top of each other, \raisebox{-0.4ex}{\includegraphics[height=1.0em]{figures/D.pdf}} visualizing color-coded skeleton for each body part with an overall score based on the calculated pose similarity by body parts, and \raisebox{-0.4ex}{\includegraphics[height=1.0em]{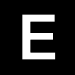}} presenting overall pose similarity through background effect changes.}
  \Description{This displays five panels capturing feedback mechanisms from a technology probe workshop, shown on a screen to participants. A) Emoticon feedback for overall pose similarity. B) Overlaid pose skeletons of different participants. C) Multiple overlaid skeletons highlighting individual discrepancies. D) Color-coded skeleton for body part accuracy. E) Overall pose similarity with an animated crowd background, summarizing performance quality.}
  \label{fig:techProbeMerge}
\end{figure*}

\subsection{Technology Probe \& Speed Dating Workshop}

\subsubsection{Technology Probe Workshop}
The technology probe workshop had two objectives: 1) to increase participants' familiarity with vision-based interactive systems and 2) to identify limitations and opportunities for integrating these technologies into a real-world dance studio \cite{hutchinson_technology_2003, zhou_dance_2021}. Given the lack of such systems in current studios and the lack of real-time vision-based systems for group dancing, this session served as a primer on the capabilities of these systems, thereby informing subsequent discussions during the Speed Dating Workshop.

\textbf{Technology Probes: }
Five technology probes were developed according to the following criteria: 1) demonstrating how vision-based systems capture and evaluate behavior, 2) low implementation cost and time, and 3) alignment with storyboard concepts presented in the Speed Dating Workshop. The first probe displayed pose similarity through discrete and value feedback \cite{gibbons_feedback_2004, raheb_hci_2016}, employing emojis based on a 5-point scale ("Excellent" to "Poor") (\autoref{fig:techProbeMerge}-\raisebox{-0.4ex}{\includegraphics[height=1.0em]{figures/A.pdf}}). The second and third probes visualized participants' own skeletons overlaid (\autoref{fig:techProbeMerge}-\raisebox{-0.4ex}{\includegraphics[height=1.0em]{figures/B.pdf}}) on top of their bodies or skeletons of participants other than themselves overlaid on top of their bodies (\autoref{fig:techProbeMerge}-\raisebox{-0.4ex}{\includegraphics[height=1.0em]{figures/C.pdf}}). The fourth probe presented pose similarity overall and by body part, using a color-coded system on the skeleton visualization (\autoref{fig:techProbeMerge}-\raisebox{-0.4ex}{\includegraphics[height=1.0em]{figures/D.pdf}}). This probe gave continuous and corrective feedback with colored skeletons \cite{zhou_visualizing_2019, zhou_syncup_2021}. The fifth probe metaphorically reflected the similarity of current poses by altering the number of people visible in the concert hall with a 5-point scale (\autoref{fig:techProbeMerge}-\raisebox{-0.4ex}{\includegraphics[height=1.0em]{figures/E.pdf}}). All probes utilized an augmented mirror interface, consistent with prior research, to focus on vision-based modalities.

We presented the technology probe using an augmented mirror primarily because we explored a vision-based system, not to exclude other information delivery modalities such as auditory or haptic. Through the technology probe, we wanted participants to experience and understand how a vision-based system captures and evaluates their movements. Thus, following previous studies\cite{zhou_movement_2022, zhou_here_2023, martinez_plasencia_through_2014}, we used an augmented mirror as it is a mature and expressive way to showcase such experiences.

\textbf{Implementation: }
All features were implemented using OpenCV\footnote{OpenCV: https://opencv.org/} in a Python environment. We employed MoveNet for multi-person pose estimation due to its ability to perform over 30 FPS on smaller devices \footnote{MoveNet, Tensorflow Hub: https://www.tensorflow.org/hub/tutorials/movenet}, which makes it suitable for real-world dance studio contexts. While methods like cosine similarity\cite{kang_dancing_2023, guo_dancevis_2022, guo_phycovis_2021} and dynamic time warping\cite{zhu_zoombatogether_2023, kim_real-time_2018, ferguson_dynamic_2014} were considered to calculate pose similarity, we chose not to utilize them. We decided that using Body-part level Pose Distance (BPD) values \cite{zhou_syncup_2021, lee_cheerup_2023} was sufficient because we wanted to 1) rapidly explore the design space before building a relatively robust concept with a high-quality movement dataset for each session participant and 2) provide an overall experience with a vision-based interactive system for evaluating dance movements. Thus, we use the same method to calculate the normalized $BPD(i,t)$ for each body vector $i$ at a given frame $t$ \cite{zhou_syncup_2021, lee_cheerup_2023}:

\begin{equation}
  BPD(i,t) = \left ( \frac{d(i,t)}{J} \right )^{\lambda} = \left ( \frac{\sum_{J}\left | \vec{v}_{j}(i,t) - \vec{v}_{R}(i,t) \right |}{J} \right )^{\lambda }
\end{equation}

where $\vec{v}_{j}(i,t)$ is the $i$-th normalized vector ($i\in\{1,2,….,14\}$) of the $j$-th person($j\in\{1,2,…J\}$), and $\vec{v}_{R}(i,t)$ is the $i$-th normalized reference vector, as same as the average of the normalized vectors of all participants. The lambda is a parameter used to set the sensitivity of the pose discrepancy. To compute the overall pose similarity, a simple addition of the BPD vector values was used for a simple implementation.

\begin{figure}[]
  \centering
  \includegraphics[width=0.5\textwidth]{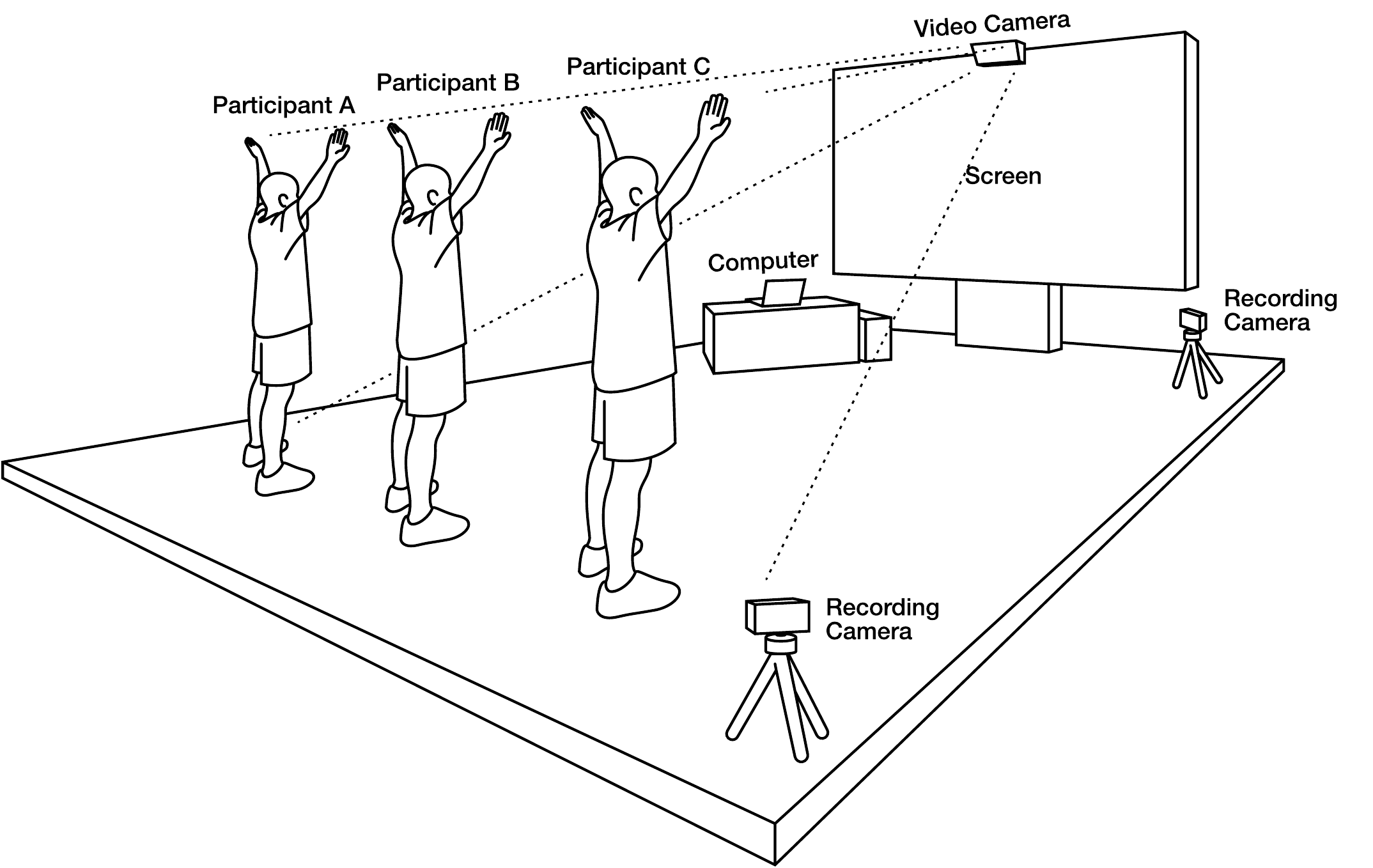}
  \caption{Setup: A large monitor was positioned at the front of the space, and a Logitech Stream camera placed on top of the monitor received the dancers' movements to provide real-time feedback. Each session was conducted with three participants, and the sessions were recorded via two GoPros located in the front and back corners of the space.}
  \Description{This oblique 3D diagram depicts the configuration used in a technology probe workshop. It shows three participants and the key equipment: a front-positioned monitor displaying real-time feedback, a Logitech Stream camera above the monitor, and two GoPro cameras capturing the session from the front and back of the space.}
  \label{fig:technologyProbeSetting}
\end{figure}

\textbf{Setup and Procedure: }
The experimental setup included an 86-inch LED TV (Samsung 86UHD-MN) to serve as the augmented mirror and a Logitech StreamCam for real-time capture within a 5m x 6m indoor space (\autoref{fig:technologyProbeSetting}). All sessions were recorded. Each probe was explored for approximately 20 minutes through free movement and set cheerleading routines. Participants did not see the skeleton visualization during the first and fifth probes; it was displayed from the second through the fourth probes. The reason for withholding the skeleton visualization in the first probe was to assess natural interaction with the system and its inherent reliability. It was displayed in subsequent probes to offer insights into how body motions were captured and evaluated. The final probe again omitted the skeleton visualization to align with storyboard concepts that did not always employ an augmented mirror.


\subsubsection{Speed Dating Workshop}
Similar to previous research \cite{zhou_here_2023}, the Speed Dating Workshop was designed to offer a rapid, cost-effective assessment of the expanded design space explored in earlier phases \cite{zimmerman_speed_2017}. The session evaluated the challenges and opportunities of various design alternatives for practical implementation in real-world dance studios. Leveraging insights from the preceding Technology Probe Workshop, participants engaged with multiple storyboards to consider the practicalities of incorporating vision-based interactive systems into the cheerleading dance studio (\autoref{fig:storyboards}). Discussions centered on participants' segmented experiences with each concept, their reliability in practical settings, their potential impact on dance studios, and avenues for combining and improving the systems.

\subsection{Participants \& Procedure}
We employed a two-tiered approach to evaluate our design alternatives~\cite{zhou_here_2023}: a Technology Probe Workshop~\cite{hutchinson_technology_2003, zhou_dance_2021} followed by a Speed Dating Workshop~\cite{zimmerman_speed_2017}. These methods afforded us both a technological examination and a user-experience-centered evaluation of vision-based interactive systems for cheerleading practice~\cite{zhou_here_2023}.

For the study, we recruited 9 participants (2M/7F) with experience in the same cheerleading crew we previously studied. The average age of the participants was 20.11 (SD=2.03), and they had been learning cheerleading as students for an average of 10.89 months. Three participants (P6, P7, and P9) had been cheerleading instructors for an additional year. In addition to cheerleading, the participants had an average of 28.67 months of experience learning group dances such as K-pop or ballet. However, their experience with computational-interactive systems such as technology probes was limited, with an average rating of 0.33 on a 4-point Likert scale (0-Never, 3-I consider myself an expert). Specifically, each participant had performed at least once with the crew and was either currently active or had been active for a minimum of one year in the past.


The study was conducted in teams of three participants and ran three times. As a token of appreciation, participants were compensated with 25,000 KRW. Each session spanned 2.5 hours and was split into two distinct activities:

\textbf{Technology Probe Workshop(30m)}: The first half-hour was allocated for the Technology Probe Workshop. Participants with limited prior exposure to computer vision-based systems participated in this section. Inspired by prior research \cite{zhou_here_2023}, the aim was to evaluate the technological feasibility and ascertain user engagement with our proposed interactive cheerleading systems.

\textbf{Speed Dating Workshop(2h)}: The following two hours were dedicated to a Speed Dating Workshop. Here, participants engaged with storyboards depicting the interactive concepts that leveraged computer vision techniques. The primary objective was to swiftly gauge user receptiveness and fine-tune the design alternatives based on real-world studio scenarios. The extended duration allowed for in-depth discussions concerning potential challenges and opportunities should these systems be deployed in their dance studios.

These workshops provided comprehensive insights into the strengths, limitations, and potential improvements for vision-based motion assessment technologies in cheerleading practice.


\subsection{Potential User Experience}
Since our overall research has focused on exploring and expanding a design space for vision-based interactive systems in group dance contexts, we summarize the results of evaluating this design space through technology and speed dating workshops. The lead author utilized thematic analysis methods to analyze and interpret what was discussed at the workshops \cite{braun_using_2006, hsieh_three_2005}.

\subsubsection{Technology Probe Workshop Results}
\mbox{}

\textit{\textbf{Initial Expectations:}}
Participants had initial expectations of the system before trying it out. Some participants expected the AI system to provide consistent and objective ratings. "\textit{Sometimes one instructor thinks it's right and another thinks it's wrong, so I think this system would be an objective and accurate assessment.} (P1)" There was also an expectation that the cheerleading choreography would track detailed body parts (hand movements, head turns, etc.) that are traditionally difficult to see with the human eye. "\textit{When we're practicing detailed movements, we're checking a lot of hand gestures, like turning our heads to each other, so I think there was an expectation that those things would be tracked.} (P2)"

\textit{\textbf{Initial Explorations:}}
Participants were initially free to test the technology's challenges and opportunities. They tested how accurately and quickly the system captured their bodies. In particular, they focused on movements that are typically utilized in cheerleading and tested how well it tracked them when they performed occlusive or vigorous movements. "\textit{I was curious about how far the skeleton was tracked, so I wanted to see if wrists and hands were tracked as well, to see how well it would recognize the body when turning around.} (P7)" They also tested the system by matching movements to each other and making small body movements to see how sensitive the system is to assess the similarity of movements. "\textit{Since the system is measuring similarity between people, I wanted to get a little bit of a baseline of what it considers to be similar.} (P7)"

\textit{\textbf{Critiques:}}
Participants criticized the vision-based technology for its lack of real-world use in a dance studio. In particular, P6 mentioned that the camera's recognition speed is slightly delayed, confusing because it doesn't match your movements and the sound of your footsteps. P2 also pointed out that the camera's field of view is too narrow to recognize multiple people. In addition, another participant pointed out that the narrow angle of view forces people to stand close to each other, and when performing vigorous movements such as cheerleading, body parts overlap, and tracking becomes difficult at that moment. "\textit{I think it would be better to use a wide-angle camera or something like that because if we're all crammed together on a small screen, our bodies might overlap.} (P7)" One participant also criticized the limitations of 2D tracking, saying that even when the system thinks the movements are the same, they may actually be slightly different. "\textit{When we actually showed it, it felt a little out of order because we weren't matching the shape (the skeleton); we were matching the behavior (the body).} (P6)"

In addition to the vision-based pose evaluation system, some participants criticized the augmented mirror and its visual effects. P6 said that in the third probe (\autoref{fig:techProbeMerge}-\raisebox{-0.4ex}{\includegraphics[height=1.0em]{figures/C.pdf}}), it was difficult to focus on other participants' skeletons and pay attention to their own movements, and P4 said that it was difficult to know if the movements were correct because the skeletons of other people were different from their actual body shapes. In addition, some participants said that in the fifth probe (\autoref{fig:techProbeMerge}-\raisebox{-0.4ex}{\includegraphics[height=1.0em]{figures/E.pdf}}), the visual effects actually covered the mirror, which had the negative effect of distracting them. "\textit{I don't really see much of a difference, and I think it's more distracting to have to watch the motion and see the background change underneath.} (P4)"

\textit{\textbf{Suggestions:}}
Participants provided several suggestions for the use of the technology in the context of real-world group dance practice. First, there were suggestions for improving the limitations of the pose tracking technology. P7 suggested that due to the limitations of the current 2D-based tracking technology, depth should also be measured, or a multi-view would be needed to evaluate cheerleading movements more rigorously. In addition, some participants wanted more control over the system. They wanted to adjust the sensitivity with which the system calculates pose similarity and set the duration (window size) for which feedback is provided if it is discrete. They felt the system's settings should differ depending on the practice context. "\textit{I think if you make the system a little more sensitive when practicing the basics initially, the psychological pressure will make you work harder.} (P9)"

There were also responses about how the provided technology could be used differently in a dance studio: P6 said that it would be suitable for demonstrating and explaining the mechanics of the joints in a cheerleading move rather than for assessment purposes. "\textit{If the instructor has a skeleton visualization and says that the joints of the body follow this route, I think it's a little bit more intuitive to show how the body should move and the mechanism of the joints and stuff like that.} (P6)"

Participants also discussed how reliable these systems would be in practice in a dance studio. P9 said that the initial experience with the system would be important. In particular, P1 said they would naturally trust it when it showed them how their body was being captured, like a skeleton visualization, or when they saw the scores change in real time. Despite this, several participants said they did not fully trust the system. P6 and P7 felt that the systems were similar to the machines that rate karaoke songs and would never fully replace human motion assessment.

\subsubsection{Speed Dating Workshop Result}
\mbox{}

\textit{\textbf{Reflections on the Design Space for General Practice Apps}}

\textit{Augmented Mirror with Visual Cues:}
Participants discussed how the visualizations provided through the augmented mirror would be experienced in actual practice. Some participants felt that the visualizations provided through the mirror were distracting and took away from the role of the mirror. "It was a bit distracting. I don't know if it's used for games and stuff like that, but other than that, I don't think it's appropriate for a training situation. (P4)" P9 said that providing real-time visual feedback should be subtle enough to be perceived by squinting. In addition, P7 believed that those who can fully absorb the information provided by the mirror during intensive movement are already skilled enough not to need such an auxiliary system. P7 also highlighted that different systems are needed at different times of practice. "I think that someone who can check themselves by looking at the visual information in the mirror at that moment while dancing is already somewhat skilled (P7)". However, university cheerleading crews usually practice for a specific performance. As such, participants noted that the closer they get to the performance (and the more experienced they thus are), the more they have to practice without mirrors and that visual aids would inherently become less applicable. "At the end of the day, the performance itself is done in a situation with no visual support, so I don't think visuals will be very helpful at some point. (P6)" P2\&P3 felt that auditory cues would be more suitable for real-time feedback.

\textit{Discrete Feedback vs Continuous Feedback:}
Participants discussed whether feedback should be provided continuously or discretely. Generally, when feedback was provided continuously, it was presented through a score (0 to 100) or color (0 to 255) to provide a more precise indication of the similarity of actions. On the other hand, discrete feedback was presented in five bins (Excellent-Good-Fair-Poor-Bad) to give a rough idea of the similarity of behaviors. P4 and P6 felt that individualized feedback was not really constructive because it seemed to say that only their movements at that moment were correct. P9 felt that the continuous feedback was more reliable and that the more precisely they could assess the similarity of the movements, the more reliable the system would be in practice, as it is vital in cheerleading to match each other's movements completely.

\textit{Feedback Timing:}
Participants also broke down the timing of when feedback can be delivered to dancers. Typically, corrective feedback is given immediately after an incorrect movement. However, there was also discussion of feedback given before or simultaneously with a movement, such as confirmatory or encouraging comments during correctly performed actions. Regarding feedback before performing a particular movement, P9 said that being told to push through a physically challenging section of the movement would fire her up. Another participant said that if she repeatedly got a certain section wrong, guiding her on what to do or shouting at her to focus on that section would make her more tense and focused (P1), states which she regarded as beneficial to her learning. In addition, one participant said it would be helpful to be told the beats one after the other, especially if they were starting the movements sequentially. "I have friends who can't count the beats, so I think it would be good to have a reference for them so that if they miss something like that, they can now count the beats with a blinking time or something like that so that they can count the beats themselves. (P7)"

\textit{Evaluation Metrics:}
The participants elaborated on the more critical information that should be conveyed to the dancers during actual cheerleading practice situations. P4 mentioned that it is not very meaningful to point out someone who made a move wrong or only to communicate the score, but it is essential to communicate how to improve the move. In addition, P7 suggested the possibility of a new dance evaluation metric, not just calculating pose similarity or temporal alignment, but something evaluated sequentially in a practice session from a macro perspective. "I think it's a little less intuitive to have a score for each body part... In the training process, it's more like 'let's match the movements!' rather than 'let's train the arms!' I think it would be nice to have a score for each training process... Some points for memorizing the choreography, some points for matching the movements, some points for detailed movement... (P7)"

\textit{\textbf{Reflections on Design Space for Group-Specific Apps}}

\textit{Group Reflection:}
The participants discussed their experiences with reflective feedback in cheerleading practice involving group rather than individual reflection. In particular, P6 and P7 mentioned that the concept of generating sounds with body movements to match the similarity of their movements through sound (\autoref{fig:storyboards}-\raisebox{-0.4ex}{\includegraphics[height=1.0em]{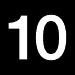}}) was similar to how they use footsteps to check the beat in actual practice. In particular, some participants hoped that this concept would encourage them to collaborate with each other during practice. "\textit{I think it would be good because it would give us a small pleasure to match the sound, and it would give us a sense of cooperation to try to match it together.} (P4\&P6)" Regarding the concept of overlaying each other's body skeletons (\autoref{fig:storyboards}-\raisebox{-0.4ex}{\includegraphics[height=1.0em]{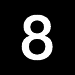}}), one participant said that it was similar to the practice of wearing colored bracelets while standing in a line to match movements in real practice. "\textit{Like we actually practice with colored bracelets, I think if we have our own skeleton color, we can recognize it a little bit better.} (P2)"

\textit{Public Feedback vs Private Feedback:}
Participants discussed receiving feedback publicly versus privately in group practice situations. P1, P5, and P6 said that they would feel stressed if they received feedback publicly in front of others that they were constantly doing poorly. On the other hand, some participants said that public feedback actually made them feel more socially tense. "\textit{I think it's really effective if you're motivated, but if you're not, I think you'll ignore private feedback.} (P1)"; "\textit{I think if we all listen to someone's feedback together, we can point out what they're doing wrong, so I think we need to know what we're doing wrong.} (P3\&P4)"

\textit{\textbf{Challenges of Interactive Systems in Dance Studios}}

\textit{Quantified Evaluation Results:}
Participants mentioned the quantitative metrics that the system provides after evaluating their behavior. P1 said that these quantitative indicators gave him confidence in the system. On the other hand, some other participants were concerned that these quantitative metrics would hurt the group atmosphere. "\textit{I'm not sure why there's a score... The whole point of using a system like this is to improve something that's a little bit wrong, to make it better, not to see who's better...} (P4)"; "\textit{I'm comfortable with vague criteria, but I think it would bother me if my friend next to me got a 75 and I only got a 74.} (P6)"; "\textit{I think cheerleading is all about making the same moves together, so I don't think we need to break down who's good and who's not.} (P7)" In this regard, P7 believes that objective metrics such as grades should be passed through the instructor to the student as much as possible.

\textit{Impact on Group Dynamics:}
Participants were concerned about the potential impact the system could have on group dynamics. Participants were concerned that the system could confuse the role of the instructor: "\textit{I think there could be cases where the instructor gives feedback that the behavior is wrong, but the students rebel because the AI says their behavior is right.} (P7)"; "\textit{I think there could be cases where the system and the instructor disagree. Since the behavior-matching process is centered around the instructor, who is also the crew's president, I think the instructor's opinion should come first.} (P1, P2)" Additionally, P8 mentioned possibly taking on an unwanted role. "\textit{In a small group context, if the system says I'm the best at something, I feel like I have to lead an exercise even if I don't want to lead the group.} (P8)"

\textit{\textbf{Opportunities for Interactive Systems in Dance Studio}}

\textit{Purpose, Role, Use:}
Participants discussed how and for what purposes such a system could be utilized within a dance studio. Some participants mentioned that such a system could be used to set common goals and could be utilized for evaluation purposes. "\textit{I feel like we're more motivated to get it right.} (P7)"; "\textit{I think we'll practice movements a few times amongst ourselves and then use the system as a final evaluation.} (P2\&P3)" Participants also mentioned the need for the system to be selected and adapted to consider the group's characteristics and the practice period. "\textit{For example, it's okay if the music is interrupted when we're trying to figure out the moves together, but at the beginning, we don't know the choreography very well, so we're taught through skeleton visualization...  I think a different system should be used for each stage of learning.} (P4)" In addition, P9 noted that the choice of system should be considered in relation to the characteristics of the group. "\textit{I think it would be a lot of fun to use this as a game, like a good-natured competition, but I think you'd have to use it carefully depending on the group's disposition.} (P9)"

P7 also mentioned that such a system would serve as a tool to assist instructors. He mentioned the process and dilemma of preparing amateur student dance crews to perform and noted that balancing the use of interactive systems would be essential. "\textit{In the end, we practice for performance, but the higher the performance quality, the harder it is to practice and the less satisfied we are with the club. Since we are amateur students, we can be divided into those who want to enjoy club activities and those who want to make a great performance. It would be better if the characteristics and diversity of these amateur groups could be considered and used appropriately.} (P7)"

\textit{Re-purposing and Suggestions:}
Participants offered other suggestions for re-purposing the system, particularly P7, noting that the concept of making sounds through the body could be utilized in choreography where the same movement is initiated on different beats. "\textit{I think if you use this system, even if you close your eyes, you can tell if you start the movement on the right beat.} (P7)" P7 also mentioned that it might be better to set a threshold for the score, showing only pass or fail depending on the context. "\textit{If the dancer crosses the threshold, the score is passed, and the dancer is simply informed of the pass, and if the dancer fails, the detailed score is given to the instructor so that they can refer to it in their practice.} (P7)"

%% file: sections/08_Discussion.tex
\section{Discussion}
\subsection{Navigating Feedback Dynamics}
Through observation and in-depth interviews conducted within the dynamic context of cheerleading, we identified 24 key insights (\autoref{tab:insightsList}). These insights pointed to a clear preference for short auditory feedback during the intensive movements of routines and specific visual feedback or explanations after the movement has concluded. These distinctions underscore the need for interactive systems that offer feedback adaptable to the physical nature of dance, its timing, and the different learning stages dancers progress through. Additionally, we identified it is important for interactive systems to adjust their feedback to suit both the immediate and reflective dance learning phases. Furthermore, our research unearthed insights related to group dynamics, highlighting the intricate balance necessary in offering personalized feedback that supports individual development while maintaining group cohesion and avoiding discomfort among dancers.

Our findings indicate that designing interactive dance systems requires sensitivity to each group's distinct cultural and organizational attributes. Cheerleaders in our study preferred avoiding public correction, stemming from the stringent, synchronized nature of their routines within a hierarchical structure. In contrast, systems like ZoombaTogether allow for public yet anonymous feedback, with studies showing that participants are motivated by the understanding that they are not alone in their challenges, rather than feeling singled out \cite{zhu_zoombatogether_2023}. This reflects Kurt Lewin's field theory \cite{lewin1942field}, which suggests behavior results from personal and environmental factors. 
Thus, future interactive dance systems should be versatile, reflecting group dance's diverse cultural and organizational settings.

Our research contributes to HCI by shedding light on a previously unexplored domain. We highlight the critical role of context in discerning user needs and illustrate how domain-specific insights can steer the development of systems that bolster group dance practice. However, we must also acknowledge the limitations of our work: we concentrated on small group dance practice during cheerleading, a genre characterized by amateur performers, hierarchical social structures, and synchronized, intense movements. 
While this focus was apt for our investigation, future research should adopt requirements for different contexts, such as larger groups or varying group atmospheres and organizational cultures, which may alter user needs substantially. More comprehensive research using methods that grasp the dynamic nature of group interaction is necessary to accurately tailor interactive system designs to the varied landscapes of group dance practice.

\subsection{Expanding and Evaluating the Design Space}

We sought to expand and understand the design space for interactive systems to support group dance learning through two methods. Our technology probe workshop highlighted the technical challenges and considerations of using RGB cameras for capturing poses during group dance practice, particularly regarding their portability and suitability for real-world applications. Contrary to prior research on augmented mirrors for improvisational dance\cite{zhou_here_2023}, issues such as delay and poor depth perception were identified as disruptive in structured group dance practice. Delays in feedback were reported to disrupt the learning experience, while a single RGB camera's limited depth perception impairs accurate movement evaluation, which is crucial in synchronized dance performance. 
Additionally, to mitigate occlusion issues among dancers---a factor less considered in traditional vision-based systems---the spatial arrangement of camera fields of view needs careful planning to ensure all dancers are clearly in sight. It is insufficient to simply consider camera location, as occlusions can occur in quite complex patterns. Solutions that address these practical challenges will enhance system tracking capabilities and preserve the integrity of group formations, improvements that will make a significant step towards effectively integrating RGB cameras in group dance settings.

In contrast, our speed dating workshop provided nuanced insights into user experiences with our 15 storyboards detailing concepts for interactive systems for group dance practice. The feedback mechanisms elicited mixed reactions, with visual cues being distracting for some, yet others saw potential benefits if these cues were subtly applied. This underscores the need for a design approach that carefully tailors feedback to support rather than interrupt the dance experience. Furthermore, the discussions on the timing and modality of feedback highlighted a preference for preventive feedback, offering guidance just before complex movements. This approach to feedback, which is relatively under-explored in dance HCI research, is a key area for future exploration. We recommend that future system designs aim to include content that prepares and guides dancers for upcoming movements.

The perceptions of quantitative score-based feedback elicited critical reflections on trust and group dynamics. While the accurate, individual objective performance measurement enhanced the system's trustworthiness, it was also reported as potentially problematic and demotivating. It was viewed as introducing a competitive element to the practice that was misaligned with the collective goal of synchronized movement inherent in cheerleading. However, the social comparisons that arise from dance cheerleading's explicit, objective goals and quantitative performance metrics cannot simply be explained by traditional social comparison theory~\cite{festinger1954theory}. Rather, the feedback highlighted concerns over creating an unnecessary competitive hierarchy that could detract from the group’s cohesion and the shared goal of synchronicity. This was particularly evident in responses indicative of upward social comparison, where lower scores made participants feel they were detracting from the group’s collective effort. This sentiment reflected the high cohesion and collectivist orientation of the groups we studied and their emphasis on the importance of unity in achieving synchronized movements.

The transition in participant attitudes towards public versus private feedback throughout the study was striking. During the initial formative study phase, participants preferred privacy in feedback to avoid psychological pressure. However, in the speed dating workshop, opinions shifted towards favoring public feedback to increase the efficiency of practice sessions and group cohesion. This significant shift in feedback preference from private to public can be attributed to an enhanced understanding of how public feedback could streamline the practice process, reinforcing group cohesion and facilitate a more collaborative learning environment \cite{zajonc1965social, cottrell1968social}. This evolution in preferences underscores the importance of adaptable feedback mechanisms that cater to the dynamic nature of group cohesion and learning needs over time.

Participants also highlighted the positive impact of systems that promote group reflection and cooperation. In the highly synchronized context of dance cheerleading, the collective goal is paramount, rendering individual competition irrelevant. Systems like sound-based group reflection were appreciated for enhancing group cohesion by encouraging collective effort toward achieving synchronized movements. This feedback underscores the importance of designing systems that provide corrective feedback and promote group reflection, ensuring that technological interventions align with the collective objectives and supportive dynamics of group dance practice.

\subsection{Design Implications} 
Our study's foray into interactive systems for group dance highlighted the utility of RGB cameras for capturing poses: they are highly portable and suitable for real-world settings. Yet, this approach unveiled challenges, such as managing user expectations and acquainting them with new technologies. This underscores the importance of an initialization session to set realistic expectations about the system’s capabilities. Furthermore, it points to the need to explore alternative pose-capturing technologies that offer portability while addressing the limitations of vision-based systems. For instance, the VoLearn system uses phone IMUs for measuring body movements and employs Dynamic Time Warping (DTW) for pose evaluation \cite{xia_volearn_2022, xia_volearn_2021}, offering a solution to vision-based systems' occlusion issues.

Regarding group dynamics, our research highlighted the critical role of feedback in influencing group dynamics, fostering either competition or cooperation. We did not initially consider "anonymity" as a design dimension, suggesting future exploration in this area is needed. Anonymity in feedback mechanisms, exemplified by systems like ZoombaTogether\cite{zhu_zoombatogether_2023}, could reduce competitive tensions, promoting a supportive practice environment. This introduces a potential design direction in the form of systems that maintain motivational feedback without fostering competition. Additionally, the concept of feedback that encourages group reflection rather than individual correction represents an area with significant potential for future HCI research that focuses on enhancing cooperative dynamics. According to collaborative learning theory, which posits that learners engage in a social and interactive process to construct knowledge through discourse and collaboration within a community \cite{dillenbourg1999you}, such group-centric feedback mechanisms could effectively support this process. The relationship between technology, students, and dance studio instructors also warrants further investigation to ensure future novel interactive systems genuinely enhance the existing learning experiences.


The application of interactive systems across various group dance contexts underscores the necessity for adaptability to meet specific genre requirements. For example, our sound-generating system designed for group reflection, as opposed to its utilization in improvisational settings like CO/DA \cite{francoise_coda_2022}, exemplifies the varied potential applications of these systems in dance. This flexibility is crucial, highlighting the importance of crafting systems catering to differing goals—from evaluation and motivation to enhancing practice. The varied feedback preferences of participants, influenced by their practice contexts, suggest a significant need for adaptive systems such as adapt2learn \cite{turakhia2021adapt2learn}. These systems could leverage educational principles like scaffolding and fading to dynamically adjust students' levels dynamically, further enriching learning experiences \cite{puntambekar2005tools}. Exploring suitable feedback modalities~\cite{chauvigne2019multi, trajkova_takes_2018} and dynamics in large groups, particularly for dances that require precise coordination, opens an expansive field for research. Developing systems to guide and synchronize movements in extensive group practice could profoundly enhance both the effectiveness and enjoyment of group dance activities, thereby expanding the reach and impact of interactive dance systems in creating supportive, engaging, and rewarding experiences.

%% file: sections/09_Limitations_FutureWork.tex
\section{Limitations \& Future Work}
One of the limitations of our approach is the expansion of the design space without the full development and deployment of a comprehensive system prototype. While technology probes provided valuable insights into potential system features and user experience in practice, they did not constitute a fully realized system capable of extensive testing in diverse group dance settings. Additionally, our participant sample was limited to cheerleading teams from a single university. This meant we focused on synchronized dance, intensive movement, and amateur-level practice in a hierarchical organizational culture. While our findings can apply to group dance genres with similar characteristics, further research is needed to understand the requirements and dynamics of group dance practices that do not share these attributes.

Future work could involve developing a comprehensive interactive system for group dance practice based on the insights gained and design concepts validated through our formative research and technology probes. This next phase would include detailed system development and extensive field deployments to rigorously test the system's effectiveness, scalability, and adaptability to various group dance contexts beyond amateur cheerleading practices. Long-term impact studies could be conducted to investigate the effects of interactive systems on group dynamics, learning motivation, and skill acquisition over time. Furthermore, developing adaptive feedback mechanisms that can tailor feedback based on the evolving needs and skill levels of individual dancers and the group could be explored to enhance the learning experience further. Lastly, future work could delve deeper into the pedagogical aspects of integrating interactive technologies into group dance practice, collaborating with dance instructors and educators to develop teaching strategies and methodologies that leverage these systems for a more holistic approach to technology-enhanced dance learning.





%% file: sections/10_Conclusion.tex
\section{Conclusion}
This paper expands on the design space of interactive systems for group dance practice by conducting user-centered studies with amateur cheerleading troupes. We identified user requirements related to feedback dynamics in group dance practice through observations and interviews. Building on these insights, we generated 15 storyboards of novel concepts in an ideation workshop and quickly evaluated the potential user experiences with these prospective systems using technology probes and speed-dating workshops. To ensure the practical applicability of our research, we envisioned a vision-based system that could be used in real-world group dance practice, prioritizing portability for amateur settings. Our findings contribute to HCI by highlighting the unique aspects and needs of group dance activities, which have been largely overlooked in previous research focusing on individual dance learning.

Our work offers valuable insights into designing interactive systems that cater to the complexities of group dynamics, synchronization, and collective learning in dance. By expanding the design dimensions to include feedback timing, privacy, and simultaneity, we provide a foundation for future research and development of technologies that enhance the group dance experience. Furthermore, our findings emphasize the importance of considering the complex group dynamics and contexts within group dance practices when designing interactive systems. Our study can inspire the further exploration of interactive systems in diverse group dance contexts, ultimately leading to more engaging, effective, and rewarding practices for dancers at all levels.